%% file: ms.tex
\documentclass[sigplan,screen]{acmart}
\usepackage[T1]{fontenc}
\usepackage[utf8]{inputenc}

\usepackage{enumitem}

\usepackage{latexsym,amsmath,amsfonts}
\usepackage{amsthm}
\usepackage{xcolor}
\usepackage{algorithm, algorithmicx}
\usepackage[noend]{algpseudocode}
\usepackage{graphicx}
\usepackage{hyperref}
\usepackage{multirow}
\usepackage{comment}
\usepackage{listings}
\usepackage{tcolorbox}
\usepackage{microtype}

\include{macros}
\usepackage{subcaption} 

\usepackage[font=small,labelfont=bf]{caption}
\usepackage[compact]{titlesec}
\usepackage{enumitem}
\renewcommand\footnotetextcopyrightpermission[1]{}



\acmPrice{15.00}

\copyrightyear{2022}
\acmYear{2022}
\setcopyright{rightsretained}
\acmConference[PLDI '22]{Proceedings of the 43rd ACM SIGPLAN
	International Conference on Programming Language Design and
	Implementation}{June 13--17, 2022}{San Diego, CA, USA}
\acmBooktitle{Proceedings of the 43rd ACM SIGPLAN International
	Conference on Programming Language Design and Implementation (PLDI
	'22), June 13--17, 2022, San Diego, CA, USA}
\acmDOI{10.1145/3519939.3523730}
\acmISBN{978-1-4503-9265-5/22/06}
                        
\begin{document}

\title{Turning Manual Concurrent Memory Reclamation into Automatic Reference Counting}

	\author{Daniel Anderson}
	\affiliation{\institution{Carnegie Mellon University}\country{Pittsburgh, PA, USA}}
	\email{dlanders@cs.cmu.edu}
	\authornote{Authors are listed in alphabetical order.}
	
	\author{Guy E. Blelloch}
	\affiliation{\institution{Carnegie Mellon University}\country{Pittsburgh, PA, USA}}
	\email{guyb@cs.cmu.edu}
	\authornotemark[1]
	
	\author{Yuanhao Wei}
	\affiliation{\institution{Carnegie Mellon University}\country{Pittsburgh, PA, USA}}
	\email{yuanhao1@cs.cmu.edu}
	\authornotemark[1]

  \date{}

\input{abstract}
\maketitle

\input{intro}
\input{background}
\input{automatic_smr}
\input{weak_ptrs}
\input{experiments}

\input{related}

\input{conclusion}

\section*{Acknowledgments}

We thank the anonymous referees for their comments and suggestions. This research was supported by NSF grants CCF-1901381, CCF-1910030, CCF-1919223, and CCF-2119069.

\bibliographystyle{ACM-Reference-Format}
\bibliography{biblio}




\end{document}

%% file: macros.tex
\definecolor{ao}{rgb}{0.0, 0.5, 0.0}

\newif\ifarxiv

\usepackage[aboveskip=2pt,belowskip=0pt]{caption}
\newcommand{\myparagraph}[1]{\smallskip\noindent\textbf{\emph{#1. }}}
\newcommand{\watomic}[1]{\texttt{weak\_atomic}}
\newcommand{\watomicfs}[1]{\texttt{weak\_atomicFS}}

\newtheorem{theorem}{Theorem}[section]

\newtheorem{definition}[theorem]{Definition}

 \newcommand{\Hao}[1]{}
\newcommand{\shepherd}[1]{#1}

\newcommand{\hide}[1]{}

\newcommand{\op}[1]{\texttt{#1}} 
\newcommand{\cfont}[1]{\mbox{\tt\bf\small #1}}

\makeatletter
\lst@Key{countblanklines}{true}[t]%
{\lstKV@SetIf{#1}\lst@ifcountblanklines}

\lst@AddToHook{OnEmptyLine}{%
        \vspace{-.05in}
	\lst@ifnumberblanklines\else%
	\lst@ifcountblanklines\else%
	\advance\c@lstnumber-\@ne\relax%
	\fi%
	\fi}
\makeatother

\makeatletter
\lst@CCPutMacro
    \lst@ProcessOther {"2A}{%
      \lst@ttfamily 
         {\raisebox{1pt}{*}}
         \textasteriskcentered}
    \@empty\z@\@empty
\makeatother


%% file: abstract.tex

\begin{abstract}
Safe memory reclamation (SMR) schemes are an essential tool for lock-free data structures and concurrent programming. However, manual SMR schemes are notoriously
difficult to apply correctly, and automatic schemes, such as reference counting, have been argued for over a decade to be too slow for practical purposes. A
recent wave of work has disproved this long-held notion and shown that reference counting can be as scalable as hazard pointers, one of the most common
manual techniques. Despite these tremendous improvements, there remains a gap of up to 2x or more in performance between these schemes and faster
manual techniques such as epoch-based reclamation (EBR).

In this work, we first advance these ideas and show that in many cases, automatic reference counting can in fact be as fast as the fastest manual SMR techniques.
We generalize our previous Concurrent Deferred Reference Counting (CDRC) algorithm
to obtain a method for converting any standard manual SMR technique into an automatic reference counting technique with a similar performance profile.
Our second contribution is extending this framework to support weak pointers, which are reference-counted pointers that automatically break pointer cycles
by not contributing to the reference count, thus addressing a common weakness in reference-counted garbage collection.

Our experiments with a C++-library implementation show that our automatic techniques perform in line with their manual counterparts, and that our weak pointer
implementation outperforms the best known atomic weak pointer library by up to an order of magnitude on high thread counts. All together, we show that the
ease of use of automatic memory management can be achieved without significant cost to practical performance or general applicability.
\end{abstract}

%% file: intro.tex

\section{Introduction}
\label{sec:intro}

Manually managing memory for concurrent programs is notoriously difficult and prone to errors.  One solution is to only work in fully garbage-collected languages but this is not always possible, and comes with its own performance problems, often yielding no control to the user.  On the other hand, manually managing memory can be challenging even for sequential programs, but the concurrent setting makes it significantly more difficult.  In particular, concurrent programs can suffer from read-reclaim races~\cite{hart2007performance} with potentially disastrous results.  For example, one thread could overwrite a location containing a pointer to an object, and then reclaim the memory for that object since it is not being referred to anymore.  Another thread executing concurrently could read the location just before it is overwritten.  It could then access the contents of the object, which by now might have been reclaimed and perhaps even been reallocated for another use, or returned to the operating system.

Over the past two decades, researchers have developed a broad set of techniques to avoid such read-reclaim races.  The goal of these techniques is to delay the destruction and reclamation on an object until it can be ensured that no thread can still access the object.  These techniques are generally referred to as \emph{safe memory reclamation} (SMR), and include approaches such as read-copy-update (RCU)~\cite{mckenney2008rcu}, epoch-based-reclamation (EBR)~\cite{fraser2004practical}, hazard-pointers~\cite{michael2004hazard}, pass-the-buck~\cite{Herlihy05}, pass-the-pointer~\cite{correia2021orcgc}, interval-based reclamation (IBR)~\cite{wen2018interval}, Hyaline~\cite{nikolaev2021snapshot}, and others~\cite{ramalhete2017brief, brown2015reclaim}.  All these approaches replace the destruction of an object with a retire operation, which delays the actual destruction until it is safe.  

The SMR approaches differ in how they ensure the reclamation and destruction is safe.  
The approaches can be partitioned broadly into two classes.
\emph{Protected-region techniques}, such as RCU, EBR, IBR, and hyaline,
protect regions of code, while \emph{protected-pointer techniques},
such as hazard-pointers, pass-the-buck, and pass-the-pointer, are based on protecting individual pointers.
The protected-region techniques tend to be faster since they only need a memory fence
on every critical region instead of every read, but require more space due to longer delays between a retire and reclamation.  Both classes of manual techniques, however, are difficult to use and can lead to subtle and hard to reproduce bugs.  As evidence, Anderson {\it et al.}~\cite{anderson2021concurrent} noted several instances where manual techniques of both kinds were applied incorrectly to concurrent data structures. 
\Hao{Perhaps we can expand on what the bugs here to address one of the reviewer's comments. For example, talking about people forgetting to call retire.}


An alternative approach for memory management in languages without built-in garbage collection (or even with) is to use reference counting.
Reference counting requires very few modifications for programmers to integrate into their code, and provides memory safety and leak freedom automatically as long as the programmer does not create reference cycles.
Reference counting still suffers from read-reclaim races\footnote{A read and increment of a reference count could race with a decrement and reclamation if zero.}, but the race can be managed by the reference counting library itself instead of by the user. 

Reference cycles can be broken with so-called weak pointers~\cite{JonesHM11}, allowing the cycles to be collected. Weak pointers do not prevent the collection of the objects they point to, but are different from raw pointers in that they provide a way of checking if the object they point to is still alive (i.e. not yet reclaimed), and if so, they can be upgraded to strong pointers.
Weak pointers can be used to store back/parent pointers in many data structures or for other pointers that would have otherwise caused a troublesome cycle.
As long as a node is not part of any strong reference cycle by the time it becomes unreachable, it will be automatically freed.

Owing to the ease of use of automatic reference counting, there has been increasing interest in concurrent (atomic) reference-counted pointers (both strong and weak), as evidenced by their inclusion in the most recent C++ standard (C++20), and many recent papers on the topic~\cite{tripp2018frc,correia2021orcgc,anderson2021concurrent}.
Early approaches~\cite{DMMS02,Herlihy05} suffered severe performance issues due to contention on the reference counts, but more recent approaches,
such as FRC~\cite{tripp2018frc}, OrcGC~\cite{correia2021orcgc}, and CDRC~\cite{anderson2021concurrent},
have been able to avoid this problem by temporarily protecting pointers without incrementing the reference counts.  However, even these recent efficient concurrent reference counting approaches  can have significant performance degradation relative to manual reclamation.   The CDRC paper reports up to a factor of two in performance degradation relative to manual collection via EBR.
The main issue is the use of protected-pointer techniques which require extra memory fences on every 
read (even if the count is not incremented).



In this paper, we show that reference counting can be nearly as fast as \emph{any} manual technique while using a similar amount of memory (in most cases),
thus showing that the ease-of-use of automatic approaches comes at no significant cost to practical performance. Our method is based on the technique of Anderson {\it et al.}~\cite{anderson2021concurrent}, who propose a novel way to combine
reference counting and hazard pointers. Unlike traditional methods which use hazard pointers to protect a block of memory from being freed, their key insight is that hazard pointers can be used to
protect \emph{the reference count itself} from being decremented. This simple insight leads to two crucial patterns. First, \emph{deferred decrements} allow increments to proceed without fear of racing with a decrement that might set the counter to zero, thus solving the read-reclaim race. Second, and critically for performance, being able to temporarily protect the reference count from decrements enables readers to safely read the managed object without fear of its destruction and without the performance penalty of incrementing the reference count.

Our insight is that the technique of Anderson {\it et al.}~\cite{anderson2021concurrent} can be generalized such that the hazard pointer scheme can be replaced with just about any standard SMR scheme to yield an automatic version of that scheme with a similar performance profile.
We apply this to three (very different) state-of-the-art manual techniques, EBR, IBR and Hyaline, to yield automatic versions of all three.
To the best of our knowledge, this is the first time reference counting has been combined with any manual technique outside of variations of hazard-pointers. The resulting algorithms are all
lock free, assuming that the SMR scheme being automated is lock free.
\Hao{Talk about some of the difficulties in generalizing CDRC.}

As a second contribution, we show how this framework can be extended even further to support lock-free atomic weak pointers that also allow safe reads without incrementing the reference count.
We use them to implement a concurrent doubly-linked-list based queue~\cite{doublelink}, and show that our implementation is several times faster \Hao{orders of magnitude faster?} than the only other lock-free atomic weak pointer that we are aware of~\cite{just19}.

A key challenge with weak pointers is supporting the upgrade to strong pointers efficiently. This requires being able to atomically increment the reference count only if it is not already zero.
This operation is typically implemented using a CAS-loop~\cite{gnulib} which takes up to $O(P)$ amortized time per process if $P$ processes perform this upgrade at the same time.
\Hao{Add citation to Pedro's sticky counter and mention that ours provides stronger properties (e.g linearizable read). I'm not sure if this goes in the related works or intro}
Instead, we show how to implement a so-called sticky counter primitive that supports an \emph{increment-if-not-zero} operation so that reading and incrementing/decrementing take only $O(1)$ time in the worst case.
We believe this sticky counter primitive can be used to improve other reference counting algorithms as well~\cite{herlihy2002rop:seejournalversion}, and we believe it has many applications outside of reference counting.





\myparagraph{Contributions} 
\begin{itemize}[leftmargin=*]
	\item We show that a wide range of manual SMR techniques can be made automatic using reference counting.
	\item We show experimentally that our automatic techniques have similar throughput and memory usage to their manual counterparts. (This represents a 2x-3x throughput improvement over existing concurrent reference counting implementations.)
	\item We show how to extend our reference counting techniques to efficiently support atomic weak pointers.
	\item To do so, we implement a theoretically and practically efficient sticky counter primitive.
	\item We show that our weak pointers significantly outperform existing weak pointers in practice.
\end{itemize}

\myparagraph{Outline} In Section~\ref{sec:background}, we introduce some important background information and we defer a broader discussion of related works to Section~\ref{sec:related}. 
Section~\ref{sec:automatic-smr} describes a general technique for making manual memory reclamation automatic. 
In Section~\ref{sec:weak-ptr}, we show how to extend our algorithms with support for weakly reference counted pointers to handle reference cycles.
An experimental evaluation of the techniques described in this paper is presented in Section~\ref{sec:exp}.
Finally, we conclude in Section~\ref{sec:conclusion}.


%% file: background.tex

\section{Preliminaries}
\label{sec:background}

\myparagraph{Model and Assumptions} We work in the concurrent shared memory model with $P$ asynchronous processes assuming sequential consistency~\cite{HS08}.  Memory barriers or memory ordering instructions are needed for weaker memory models, and are included in our C++ implementations.  We use the standard definitions of wait-free, lock-free and linearizability~\cite{HS08}.
Essentially, lock-freedom guarantees that some process makes progress, while wait-freedom guarantees that every process makes progress.
Roughly speaking, linearizability means that each operation appears to take effect atomically at some point during its execution interval.
Beyond reads and writes, we assume the existence of three atomic read-modify-write primitives: \cfont{compare\_and\_swap} (CAS), \cfont{fetch\_and\_store} (FAS), and \cfont{fetch\_and\_add} (FAA).  All three instructions are supported by modern processors.

\myparagraph{Manual SMR} Most manual SMR schemes have similar interfaces built around a common set of operations. These operations include:
\begin{itemize}[leftmargin=*]
\item \cfont{retire}$(x)$: Indicate that an allocated object $x$ is no longer reachable by the program, i.e., that it is safe to delete after all readers currently reading it are finished.
\item \cfont{eject}$()$: Returns a previously retired object that is now safe to delete. The caller should then free this object.
\end{itemize}
The retire operation is the critical one; it is what replaces completely manual memory management (explicit freeing). A retire operation is essentially a ``delayed free''. Rather than being freed immediately, the object is freed once any lingering readers have finished with it. The eject operation is optional and is often performed implicitly by retire, but separating the two can allow the programmer greater control over exactly when or how memory is freed.

The difference between \emph{protected-pointer} and \emph{protected-region} techniques is in how they determine when the lingering readers have finished with a retired object, making it safe to free. Protected-region techniques implement the following pair of operations:
\begin{itemize}[leftmargin=*]
\item \cfont{begin\_critical\_section}$()$: Indicate the beginning of a read critical section.
\item \cfont{end\_critical\_section}$()$: Indicate the end of the current read critical section.
\end{itemize}
For correctness, all reads of objects that are protected by the SMR scheme must be performed while inside a read critical section. A retire operation is then able to deduce that a retired object $x$ is safe to eject once all critical sections that were active at the time of its retirement have ended.

Protected-pointer techniques use the following operations instead:
\begin{itemize}[leftmargin=*]
\item \cfont{acquire}$(m)$: Indicate the intention to read the contents of a shared pointer located at the memory location $m$, and return the current value of the shared pointer.
\item \cfont{release}$(p)$: Indicate that the pointer obtained from a shared location by \cfont{acquire} is no longer being read.
\end{itemize}
All reads of objects that are protected by the SMR scheme must be done so via an acquire operation, and ended by a corresponding release operation. A retire operation is then able to safely deduce that a retired object $x$ is safe to eject once all active acquires of it at the time of its retirement have been released.
Note that in many protected pointer schemes such as hazard-pointer and pass-the-buck, the acquire operation can fail, forcing the program to retry or take a data structure specific fallback plan.

The key difference between protected-pointer and protected-region techniques is that protected-region techniques prevent \emph{all} objects from being ejected during their read critical sections, while protected-pointer techniques are more granular and only protect the objects actually being read. Protected-region techniques are therefore usually faster since they require less bookkeeping, but accumulate more garbage because they overprotect objects from being ejected.

\myparagraph{CDRC} The key idea behind CDRC is to combine manual SMR and reference counting by using hazard pointers (a protected-pointer technique) to defer reference-count decrements until they no longer race with increments. Essentially, instead of protecting an object from being freed, an acquire operation protects an object's reference count from being decremented until a corresponding release is issued, and a retire operation issues a delayed decrement, which is performed by an eject at a later time when it is not protected by an active acquire.

To achieve this, Anderson et al.~\cite{anderson2021concurrent} introduce an interface called acquire-retire, which exposes the same four operations as protected-pointer schemes: \op{acquire}, \op{release}, \op{retire}, and \op{eject}, but generalizes hazard pointers by allowing a pointer to be retired multiple times, which is not allowed by traditional hazard pointers.
This additional feature is important because each retire corresponds to a delayed decrement, and there could be multiple of those on the same pointer.

This interface allows them to implemented reference counted pointers as follows:
\begin{itemize}[leftmargin=*]
\item To copy a shared reference-counted pointer, it is first \emph{acquired} to protect the reference count from being decremented below one (which would destroy the object and create a race). The reference count is then incremented, the protection \emph{released}, and a new copy of the pointer is safely returned.
\item To overwrite a shared reference-counted pointer with a new value, the reference count of the desired value is incremented, and the previous value is replaced via an atomic exchange (i.e. \op{fetch\_and\_store}). The previous value is then \emph{retired}, which issues a deferred decrement to its reference count, which will be applied by a later \emph{eject}, but only once it is no longer protected by a corresponding acquire.
\end{itemize}

\noindent Using this idea, CDRC implements a C++ library containing three reference counted pointer types: \op{atomic\_shared\_ptr}, \op{shared\_ptr}, and \op{snapshot\_ptr}.
Applying this library just involves replacing raw pointers with one of these three smart pointer types.
The interface they support is closely modeled after \op{atomic<shared\_ptr>} and \op{shared\_ptr} from the C++ standard.
\op{atomic\_shared\_ptr} supports arbitrary concurrent accesses, but is the most expensive to use. \op{shared\_ptr} supports everything except read-write races. \op{snapshot\_ptr} is the most efficient because it avoids incrementing reference counts in common case, but cannot be shared between threads.


Figure~\ref{fig:rc-exp} shows a code snippet from Natarajan-Mittal's BST~\cite{natarajan2014fast} using both manual SMR and CDRC.
It shows that manually calling retire sometimes adds non-trivial code.
For example, all the code between lines~\ref{line:loop-start} and~\ref{line:loop-end} can be avoided with CDRC.
This loop is responsible for retiring all the nodes removed by the pointer swing on line~\ref{line:remove-nodes}.
This loop is easy to forget because in the common (sequential) case, each pointer swing only removed one internal node, and this bug has appeared in the artifacts of several published papers~\cite{wen2018interval,nikolaev2020universal,david2015asynchronized,friedman2020nvtraverse,correia2021orcgc}.
Therefore, reference counting techniques like CDRC are often easier to use and less error-prone.

%
%
%
%
%
%


\begin{figure}
\centering
  \begin{subfigure}{0.49\textwidth}
\begin{lstlisting}[basicstyle=\scriptsize\ttfamily,linewidth=\columnwidth,xleftmargin=5.0ex,numbers=left]
class Node { K key; atomic<Node*> left, right; };
class SeekRecord { Node *ancestor, *successor, *parent, *leaf; };
thread_local SeekRecord seekRecord;

void cleanup() {  // helper function called by remove()
  Node* ancestor = seekRecord.ancestor;
  Node* successor = seekRecord.successor;
  ...
  /* Update the left child of ancestor to point to sibling */
  if(ancestor->left.compare_and_swap(successor, sibling)) {  @\label{line:remove-nodes}@
    /* retire nodes on path from successor to sibling */
    for(Node* n = successor; n != subling;) { @\label{line:loop-start}@
      Node* tmp = n;
      if(getFlag(n->left)) {
        retire(n->left);
        n = n->right;
      } else {
        retire(n->right);
        n = n->left; }
      retire(tmp); } @\label{line:loop-end}@
    return true; 
  } else return false; }
\end{lstlisting}
  \caption{Manual SMR} \label{fig:manual-smr}
  \end{subfigure} \hfill


  \begin{subfigure}{0.49\textwidth}
\begin{lstlisting}[basicstyle=\scriptsize\ttfamily,linewidth=\columnwidth,xleftmargin=5.0ex,numbers=left, firstnumber=23]
class Node { K key; atomic_shared_ptr<Node> left, right; };
class SeekRecord { 
  snapshot_ptr<Node> ancestor, successor, parent, leaf; };
thread_local SeekRecord seekRecord;

void cleanup() {  // helper function called by remove()
  snapshot_ptr<Node>& ancestor = seekRecord.ancestor;
  snapshot_ptr<Node>& successor = seekRecord.successor;
  ...
  /* Update the left child of ancestor to point to sibling */
  return ancestor->left.compare_and_swap(successor, sibling)); }
    \end{lstlisting}\hfill
    \vspace{-0.95\baselineskip}
    \caption{Reference Counting} \label{fig:our-library}
  \end{subfigure}

  \vspace{0.5\baselineskip}
  \caption{\shepherd{Code snippet from Natarajan-Mittal's BST~\cite{natarajan2014fast} using (a) manual SMR and (b) reference counting (C++-like pseudocode).}}
 \label{fig:rc-exp}
\end{figure}

%% file: automatic_smr.tex

\section{Making Manual SMR Automatic}
\label{sec:automatic-smr}

In this section, we describe how to make manual SMR automatic by combining it with reference counting.  
As mentioned in Section~\ref{sec:intro}, our approach extends the concurrent reference counting algorithm (CDRC) of Anderson et al.~\cite{anderson2021concurrent} which uses a hazard-pointer-like technique (called acquire-retire) to delay reference count decrements until they no longer race with increments.
Our insight is that this approach would work for virtually any manual SMR technique, not just hazard-pointers.
\shepherd{Note that the process of converting from manual to automatic SMR is not itself automatic, but we present an easy-to-apply framework and show several examples of how to use it.}

To generalize CDRC, we first generalize their acquire-retire interface and then show that this generalized interface can be implemented from a wide range of manual techniques.
Then we show how to implement concurrent reference counting using this generalized interface.
The final step mostly follows CDRC~\cite{anderson2021concurrent}, but with some important differences in the implementation of snapshot pointers.


\begin{figure}
	\small
	\begin{lstlisting}[basicstyle=\scriptsize\ttfamily,linewidth=.99\columnwidth, xleftmargin=5.0ex,numbers=left]
class AcquireRetire<T> {
	// Allocate object of type T
	Function alloc(): T*
	
	// Delays destructing ptr
	Function retire(T* ptr): void
	
	// Returns a previously retired pointer
	// that is no longer protected.
	Function eject(): optional<T*>
	
	Function begin_critical_section(): void
	Function end_critical_section(): void
	
	// Reads a pointer from shared memory and protects it.
	// Can only protect one pointer at a time.
	Function acquire(T** ptraddr): pair<T*, Guard>
	
	// Reads a pointer from shared memory and tries to protect it
	// Can fail and return @$\bot$@.
	Function try_acquire(T** ptraddr): optional<pair<T*, Guard>>
	
	// Releases protection
	Function release(Guard guard): T*  };
	\end{lstlisting}
	\caption{Generalized acquire-retire interface.}
	\label{fig:new-ar-interface}
\end{figure}

\subsection{Generalized Acquire-Retire Interface}

Our generalized acquire-retire interface shown in Figure~\ref{fig:new-ar-interface} has several advantages over the original. 
The original interface is well-suited for capturing protected-pointer SMR techniques (because \texttt{acquire} protects a specific pointer), but not for capturing other types of SMR techniques.
We added three new methods to the interface to make it more general: \op{alloc}, \op{begin\_critical\_section}, and \op{end\_critical\_section}.
The latter two are important for protected-region techniques.
\Hao{define critical section here and emphasize that it has nothing to do with mutex.}
Adding \op{alloc} to the interface is important for techniques like IBR and HE, which require customized memory allocation.
In particular, IBR and HE tag each object with a birth timestamp on allocation.

Beyond generality, another benefit of the interface in Figure~\ref{fig:new-ar-interface} is that it gives us a clean way of implementing snapshot pointers.
In CDRC, supporting snapshot pointers requires reaching into the internals of their acquire-retire implementation.
So unlike the rest of their reference counting algorithm, their algorithm of snapshot pointers only works for their specific implementation of acquire-retire.
We fix this problem by breaking their \op{acquire} into two operations, an \op{acquire} and a \op{try\_acquire}.
Both operations return a pointer as well as a guard variable that protects the pointer.
The pointer can be unprotected at any point by passing the guard variable to \op{release}.
In HP and HE, this guard variable would be a pointer to the announcement slot that protects the pointer. \op{acquire} can only protect one pointer at a time, so the user must alternate between calling \op{acquire} and \op{release}. \op{try\_acquire} on the other hand can protect multiple pointers with different guards. However \op{try\_acquire} may fail and return $\bot$ if it runs out of guards (e.g. running out of hazard-pointers).
We use \op{try\_acquire} to implement \op{snapshot\_ptr}s in a black box manner in Section~\ref{sec:ref-cnt}.

Lastly, just like in the original acquire-retire interface, the \op{retire} operation in Figure~\ref{fig:new-ar-interface} takes as input a pointer which will in the future be returned by an \op{eject} operation when it is no longer protected.

\subsection{Implementing Generalized Acquire-Retire}
\label{sec:gen-acq-ret-impl}

\input{acquire-retire-impl}

This new acquire-retire interface can be easily implemented from almost any manual SMR technique.
Figures~\ref{fig:ar-ebr} and~\ref{fig:ar-ibr} show implementations from EBR and IBR, respectively.
In this section, we will discuss some general patterns in these implementations.
Most manual SMR algorithms combine the functionality of \op{retire} and \op{eject} into a single \op{retire} operation, but this is easy to break up into two operations.
A more important difference is that manual SMR has only been used, until now, to delay freeing objects. 
So instead of returning retired pointers to the user, their \op{retire} function calls \op{free} on pointers that are no longer protected.
We require pointers to be returned to the user because our \op{retire} can be used to delay arbitrary operations on the pointer, for example decrementing the pointer's reference count.
In our implementation of weak pointers in Section~\ref{sec:weak-ptr}, we use three instances of \op{AcquireRetire}, each delaying a different type of operation.

Another important reason for having \op{eject} return a pointer instead of directly applying the delayed operation is to prevent \op{eject} from recursively calling itself.
For example, if the delayed operation is a reference count decrement, then this might trigger recursive reference count decrements, which might lead to recursive calls to \op{eject}.
The \op{eject} operation is not guaranteed to behave correctly if called recursively, so we disallow this possibility by not applying the delayed operation inside the \op{eject}.
The final difference between our \op{retire} and the one supported by existing SMR techniques is that we allow a pointer to be retired any number of times before it is ejected a single time.
Luckily, most SMR algorithms work properly in this kind of situation even though they were not designed with it in mind.
\shepherd{Protected-pointer approaches sometimes need to be modified to keep track of the number of times a pointer is \op{retire}d and \op{acquire}d.
\op{eject} also has to be modified so that it returns only the pointers that have been \op{retire}d more times than \op{acquire}d. No such modifications are needed for protected region approaches.}

Next, we focus on how to implement \op{acquire}, \op{try\_acquire}, and \op{release}.
For protected-region SMR techniques like EBR, and Hyaline, these operations are trivial to implement because the critical section on its own is enough to protect all the pointers returned by \op{acquire}.
So \op{acquire} and \op{try\_acquire} simply load the pointer and \op{release} is a no-op.
For protected-pointer approaches like HP and PTB, \op{try\_acquire} has to look for an empty announcement slot to act as the guard.
If all announcement slots are in use, then \op{try\_acquire} fails, returning $\bot$.
For \op{acquire}, we reserve a special guard / announcement slot that cannot be used by \op{try\_acquire}.
This ensures that \op{acquire} always succeeds but it means that only one pointer can be protected by \op{acquire} at a time.

Finally, the operations for beginning and ending a critical section are implement the exact same way as in the corresponding manual SMR technique.
So for EBR, they would just announce and unannounce an epoch, and for protected-pointer approaches, they would be no-ops.

\subsection{Defining Correctness}

Just like with the original acquire-retire interface, the tricky part of defining correctness for the generalized version is handling the case where a pointer gets retired multiple times before any copy gets ejected.
Fortunately, we can use the original correctness definition with just some small modifications.
The idea behind the original definition is to map acquires to retires and ejects to retires such that if an acquire and an eject get mapped to the same retire, then the acquire must be inactive by the time the eject is executed.
This formalizes the intuition that a pointer can only be returned by eject if it is not protected by any active acquire.
We begin by defining what it means for an acquire to be \emph{active}.

\begin{definition}[active vs. inactive acquires]
\label{def:active-acquire}
We say that an \op{acquire} or a successful \op{try\_acquire} is \emph{active} between when it was invoked and when the guard it returns is passed to \op{release}.
After its guard is released, we say it is \emph{inactive}.
\end{definition}

\noindent Our acquire-retire interface imposes some restrictions on how it can be used. These restrictions are captured in the following definition of \emph{proper executions}.

\begin{definition}[proper execution]
\label{def:proper-usage}
	We say that a concurrent execution involving acquire-retire operations is \emph{proper} if (1) each active acquire is contained in a critical section, (2) each guard returned by \op{acquire} or \op{try\_acquire} is passed to \op{release} at most once, and (3) a process cannot call \op{acquire} while its previous \op{acquire} is still active.
\end{definition}

\noindent The first property in Definition~\ref{def:proper-usage} is easy to ensure by beginning a critical section before any calls to acquire and making sure all acquires are inactive before ending the critical section. The third property just says that acquire can only be used to protect a single pointer at a time.
Now we are ready to formally define the sequential specifications of acquire-retire.

\begin{definition}[acquire-retire]
\label{def:acquire-retire}
Any proper, concurrent execution can be linearized to a sequential history with the following guarantees: 
\begin{itemize}[leftmargin=*]
	\item Successful \op{try\_acquire(pptr)} and \op{acquire(pptr)} operations return the pointer currently stored in $*pptr$.
	\item Let $f$ be a function that maps each \op{acquire} returning $p$ and each successful \op{try\_acquire} returning $p$ to either a later \op{retire(p)} or $\bot$. Let $g$ be an injective (one-to one) function that maps each \op{eject} returning $p$ to an earlier \op{retire(p)}. For all $f$, there is a $g$ such that whenever $f(A) = g(E)$, the \op{acquire} or \op{try\_acquire} $A$ is inactive by the time \op{eject} $E$ is executed.
\end{itemize}
\end{definition}

 \begin{figure}
 	\small
 	\begin{lstlisting}[basicstyle=\scriptsize\ttfamily,linewidth=.99\columnwidth, xleftmargin=5.0ex,numbers=left]
class snapshot_ptr<T> { T* ptr; optional<Guard> guard; };

AcquireRetire<T> ar;

snapshot_ptr<T> atomic_shared_ptr<T>::get_snapshot() {
	auto ptr, guard = ar.try_acquire(addressof(this->ptr));
	if(guard != @$\bot$@) return snapshot_ptr<T>(ptr, guard);
	ptr, guard = ar.acquire(addressof(this->ptr));
	increment(ptr); // increment reference count
	ar.release(guard);
	return snapshot_ptr<T>(ptr, @$\bot$@); }

void snapshot_ptr<T>::release() {
	if(this->guard != @$\bot$@) ar.release(this->guard);
	else decrement(this->ptr); }

void begin_critical_section() { ar.begin_critical_section(); }
void end_critical_section() { ar.end_critical_section(); }
 	\end{lstlisting}
 	\caption{Implementing snapshot pointers using the generalized acquire-retire interface from Figure~\ref{fig:new-ar-interface}.}
 	\label{fig:snapshot-ptr-impl}
 \end{figure}
 
\subsection{Concurrent Reference Counting}
\label{sec:ref-cnt}

Using the generalized acquire-retire interface, we can implement concurrent reference counting in much the same way as CDRC.
The main difference is in our implementation of \op{snapshot\_ptr}s shown in Figure~\ref{fig:snapshot-ptr-impl}.
The code for the other two reference counted pointer types, \op{atomic\_shared\_ptr} and \op{shared\_ptr}, remains the same except for some minor updates to use the new acquire-retire interface.

\Hao{reword the next few sentences} To support \op{snapshot\_ptr}s, we implement a \op{get\_snapshot} operation which creates a \op{snapshot\_ptr} by loading an atomic shared pointer, and a \op{release} operation which destructs a \op{snapshot\_ptr}.
\op{get\_snapshot} first tries to take the fast path which consists of protecting the pointer with just a \op{try\_acquire}.
If this \op{try\_acquire} fails, then it reverts to the slow path which consists of protecting the pointer using an \op{acquire}, then incrementing the reference count of the pointer, and then releasing the previous acquire since the pointer is now protected by the incremented reference count.
In the slow path, \op{get\_snapshot} then constructs and returns a \op{snapshot\_ptr} with its \op{guard} field set to $\bot$ to indicate that the slow path was taken.
A \op{snapshot\_ptr}'s destructor calls \op{ar.release()} if it was constructed via the fast path and \op{decrement} otherwise.
As long as a process does not hold onto too many \op{snapshot\_ptr}s, \op{get\_snapshot} will always take the fast path and not perform any reference count updates.
This is why \op{snapshot\_ptr} can be significantly cheaper than \op{shared\_ptr}s.

This is different from Anderson et al.'s \op{get\_snapshot} implementation which only works for their specific acquire-retire implementation based on hazard-pointers.
In their algorithm, \op{get\_snapshot} first looks for an empty announcement location and if all of them are taken, it evicts one of the announcement hazard pointers and increments the reference count of the evicted pointer to ensure that it stays protected.
Then \op{get\_snapshot} uses the newly emptied announcement location to protect the pointer it reads.

Another difference from Anderson et al.'s implementation is that we require all racy\footnote{Two operations are said to \emph{race} if they both access the same atomic shared pointer and one of them is a write.} reads and writes on atomic shared pointers as well as all snapshot pointer lifetimes to be contained in a critical section.
This requirement means that we cannot hold onto any snapshot pointers outside of a critical section.
When applying our reference counting algorithm to a concurrent data structure, this requirement can be satisfied by calling begin and end critical section at the start and end of each data structure operation, respectively.


%% file: acquire-retire-impl.tex

%
%
%
%

\begin{figure}
	\small
	\begin{lstlisting}[basicstyle=\scriptsize\ttfamily,linewidth=.99\columnwidth, xleftmargin=5.0ex,numbers=left]
class AcquireRetireEBR<T> {
	using Guard = void; // empty type, never used
	using Epoch = int;
	Epoch emptyann = INT_MAX;
	Epoch ann[P]; // initialized to emptyann
	Epoch curEpoch = 0;
	thread_local List<pair<T*, Epoch>> retired;
	
	T* alloc() { return new T(); }
	void begin_critical_section() { ann[pid] = curEpoch; }
	void end_critical_section() { ann[pid] = emptyann; }
	void release(Guard guard) {}
	
	pair<T*, Guard> acquire(T** ptraddr) { 
	  return [*ptraddr, void]; }
	
	optional<pair<T*, Guard>> try_acquire(T** ptraddr) { 
	  return [*ptraddr, void>]; }

	// retire + eject implemented as in Figure 2 of@~\cite{wen2018interval}@ };
	\end{lstlisting}
	\caption{\shepherd{Generalized acquire-retire implemented with epoch-based-reclamation. We assume each process knows their process id $pid$.}}
	\label{fig:ar-ebr}
\end{figure}

\begin{figure}
	\small
	\begin{lstlisting}[basicstyle=\scriptsize\ttfamily,linewidth=.99\columnwidth, xleftmargin=5.0ex,numbers=left]
class AcquireRetireIBR<T> {
	using Guard = void; // empty type, never used
	using Epoch = int;
	Epoch emptyann = INT_MAX;
	Epoch beginAnn[P], endAnn[P]; // initialized to emptyann
	Epoch curEpoch = 0;
	thread_local Epoch prev_epoch = emptyann;
	thread_local int counter = 0;
	
	void begin_critical_section() { 
	  beginAnn[pid] = endAnn[pid] = prev_epoch = curEpoch; }
	void end_critical_section() { 
	  beginAnn[pid] = endAnn[pid] = emptyAnn; }
	void release(Guard guard) {}
	class Tagged<T> { Epoch birthEpoch; T t; };
	
	T* alloc() { 
		Tagged<T>* taggedObj = new Tagged<T>();
		taggedObj->birthEpoch = curEpoch;
		if(counter++ % epoch_freq == 0) curEpoch.fetch_add(1);
		return addressof(taggedObj->t); }
	
	pair<T*, Guard> acquire(T** ptraddr) { 
		while(true) {
			T* ptr = *ptraddr;
			Epoch cur_epoch = curEpoch;
			if(prev_epoch == cur_epoch) return [ptr, void];
			else endAnn[pid] = prev_epoch = cur_epoch; } }
	
	optional<pair<T*, Guard>> try_acquire(T** ptraddr) { 
		return acquire(ptraddr); }
	
	// retire + eject implemented as in@~\cite{wen2018interval}@ };
	\end{lstlisting}
	\caption{\shepherd{Generalized acquire-retire implemented with interval-based-reclamation (specifically, 2GEIBR).}}
	\label{fig:ar-ibr}
\end{figure}

%% file: weak_ptrs.tex

\section{Weak Pointers}
\label{sec:weak-ptr}

The second classical drawback of reference counting is its inability to clean up garbage that contains cyclic references. A common approach to mitigate this issue at the library level is to include a ``weak pointer'' type. Weak pointers complement shared pointers (or ``strong pointers'') by holding a reference to a shared object without contributing to the reference count. If the reference count of the managed object reaches zero, it is destroyed, despite any weak pointers that may have a reference to it.

The advantage of weak pointers over raw pointers is that, unlike raw pointers, which are unsafe to dereference if they might point to garbage (an already freed object), weak pointers can tell whether they point to a managed object that has already been destroyed. This is usually achieved by storing a second reference count that counts the number of weak pointers to the managed object. When the (strong) reference count reaches zero, the managed object is destroyed, but the control data containing the reference counts is kept intact until both the strong and weak reference counts reach zero. This allows weak pointers to safely detect when the managed object is alive by checking that the strong reference count is non-zero.

The C++ standard library includes support for weak pointers, and, as of C++20, support for atomic weak pointers. However, currently the only standard library implementation of atomic weak pointers is Microsoft's STL~\cite{microsoftSTL}, and it is lock-based. We know of one commercial implementation in the just::thread library~\cite{just19}. We describe how our approach can be extended to support weak pointers with the same properties and performance as the original approach. \Hao{Possible rewording: We describe how our approach can be extended to support lock-free weak pointers without affecting the performance and properties of strong pointers.}

\subsection{Library interface}

\begin{figure*}[t]
\includegraphics[width=0.70\textwidth]{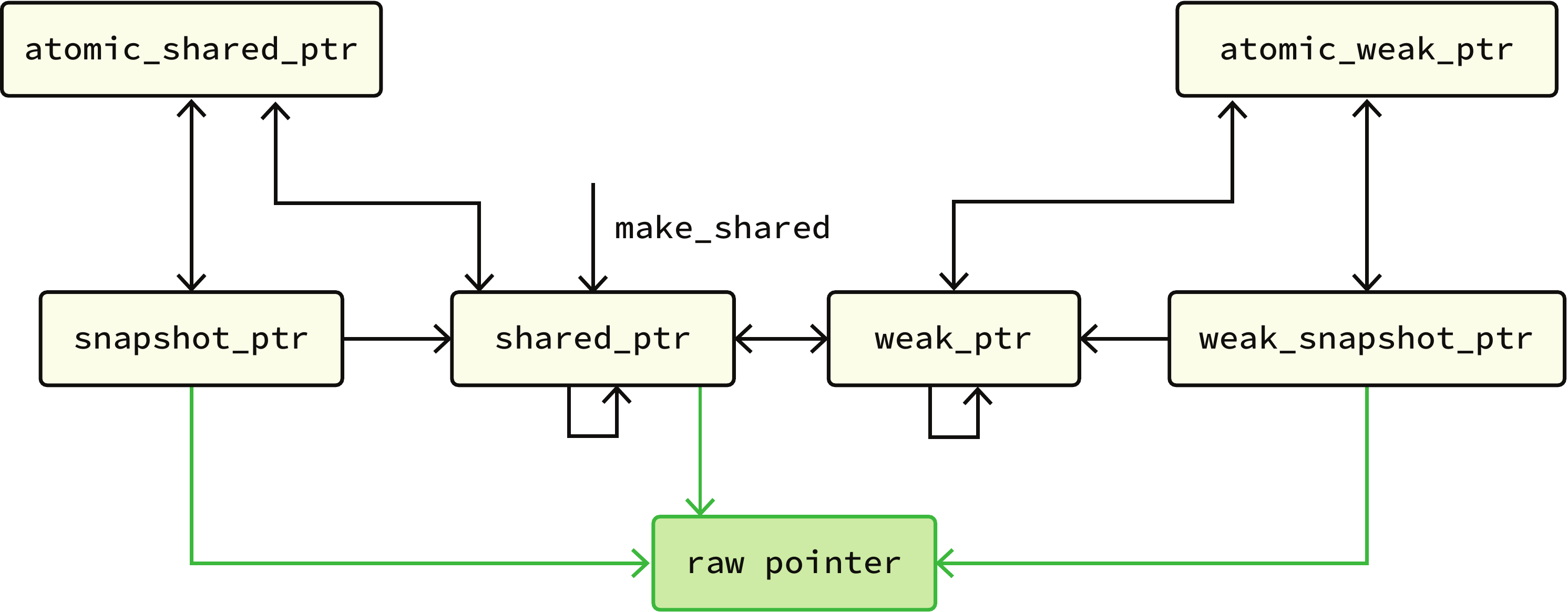}
\caption{The managed pointer types in our library. Arrows between types denote that it is possible to store/load one type in/from the other, or that it is possible to convert from one type to the other. The three types \texttt{snapshot\_ptr}, \texttt{shared\_ptr}, and \texttt{weak\_snapshot\_ptr} can be safely dereferenced/converted into raw pointers.}\label{fig:type-diagram}
\end{figure*}

We add the following types to the reference-counted pointer library. The relationship between them is depicted in Figure~\ref{fig:type-diagram}.

\begin{itemize}[leftmargin=*]
\item \textbf{atomic\_weak\_ptr}: Analogous to \texttt{atomic\_shared\_ptr}, an \texttt{atomic\_weak\_ptr} facilitates atomically loading, storing, and CASing a \texttt{weak\_ptr} into a shared mutable location. In addition to \texttt{load}, it also supports a \texttt{get\_snapshot} method, which grants safe local access to the managed object without modifying the reference count.

\item \textbf{weak\_ptr}: A \texttt{weak\_ptr} is modeled after C++'s standard weak pointer. Unlike \texttt{shared\_ptr}, a \texttt{weak\_ptr} cannot be directly dereferenced. To access the managed object, the \texttt{weak\_ptr} must be upgraded to a \texttt{shared\_ptr}. If the managed object has \emph{expired}, the obtained \texttt{shared\_ptr} will be null to indicate this.

\item \textbf{weak\_snapshot\_ptr}: A \texttt{weak\_snapshot\_ptr} provides safe access to the object managed by the \texttt{atomic\_weak\_ptr} as of the time it was created, even if the reference count of the managed object reaches zero during its lifetime. Creating and reading a \texttt{weak\_snapshot\_ptr} does not incur a modification to the reference count. A \texttt{weak\_snapshot\_ptr} will be null if the managed object has expired at the time of its creation.

\end{itemize}

\noindent The subtle difference between a \texttt{weak\_snapshot\_ptr} and a  \texttt{snapshot\_ptr} is that a \texttt{snapshot\_ptr} guarantees that the managed object doesn't expire (has reference count at least one) throughout its lifetime, while a \texttt{weak\_snapshot\_ptr} only guarantees that the managed object is safely readable, though it may expire (reach reference count zero) during the lifetime of the snapshot.

We first describe the main primitives needed to implement deferred reference counting with weak pointers. We then describe how to support the main operations on the various weak pointer types in our library.

\subsection{Managing the managed object}

First, to implement weak pointers, each managed object is augmented with a second reference count. We distinguish between the original (strong) reference count and the new (weak) reference count. When the strong reference count reaches zero, the managed object is ready to be destroyed. However, the control data attached to the managed object (the reference counts plus any extra scheme-specific metadata) cannot be destroyed and freed yet, because there might still exist weak pointers that attempt to access those fields. Only once both the strong and weak reference counters hit zero can the entire control block (the managed object plus the control data) be freed. To correctly detect when both counters hit zero in the presence of concurrent updates, we use the standard trick~\cite{gnulib,microsoftSTL} of storing
\begin{equation*}
\text{weak\_cnt} = \#\text{weak refs} + \begin{cases}
1 & \text{if \#strong refs } > 0 \\
0 & \text{otherwise}.
\end{cases}
\end{equation*}
When the strong count hits zero, it can destroy the managed object and decrement one from the weak count. To be precise, this destruction and corresponding decrement must be delayed in the presence of weak pointers. We will discuss this in Section~\ref{sec:weak_ptr_primitives}. When the weak count hits zero, the entire control block is ready to be freed immediately.

In the strong-only setting, the reference count will only ever be incremented when there already exists at least one reference, and hence the increment can always be performed with a fetch-and-add operation. In the weak setting, however, it is possible that a weak pointer points to a managed object whose strong reference count could be decremented to zero at any moment. Attempting to increment the strong reference count with a fetch-and-add could therefore result in incrementing the counter from zero, thus resurrecting a dead object. Our algorithms therefore require an \emph{increment-if-not-zero} operation, which can return false if the reference count is zero, and hence should not be incremented.

The increment-if-not-zero operation is traditionally implemented as a simple CAS loop, which continuously attempts to add one to reference count as long as it is not zero, or returns false otherwise. This results in the increment having lock free but not wait free progress. In the next section, we describe a simple, but to the best of our knowledge, novel implementation of a constant-time wait-free counter that supports the increment-if-not-zero operation. This data structure in general is sometimes referred to as a sticky counter. Specifically, our data structure implements an atomic counter that supports increment-if-not-zero, decrement, and load, all in constant time using single-word atomic instructions.

\subsection{Wait-free increment-if-not-zero} Our algorithm can implement a $b$-bit wait-free reference counter using $b+2$ bits, that is, we use two bits for bookkeeping purposes. For example, if using $32$-bit integers to store the reference count, we can handle up to $2^{30}$ references to an object. The main idea is simple, we use the highest bit of the reference counter to indicate whether the reference count is zero. Any bit pattern in which the highest bit is set is interpreted as zero, and otherwise is not. Note importantly, that this means that the stored value being zero is not interpreted as the reference count being zero! The implementation is described below and depicted in Figure~\ref{fig:sticky-counter}. This technique of using the high bits to store a flag above a counter is similar to that of Correia and Ramalhete~\cite{correia2018strong} who implement reader-writer locks that store a count of the number of shared readers. Our technique generalizes theirs by allowing constant-time linearizable reads of the counter.

\myparagraph{Increment} Since the presence of the high bit indicates whether the counter is zero, the increment operation can just perform a fetch-and-add operation, and check whether the result has the high bit set. If so, it returns false.

\myparagraph{Decrement} The decrement operation should decrement the reference count and return true if the reference count was brought to zero, or false otherwise. To decrement the counter, the algorithm uses a fetch-and-add and checks whether the counter hits zero. If it does, it must attempt to set the high bit to indicate this. This is done with a CAS. Note that if the CAS fails, it must be the case that an increment occurred that brought the counter back up from zero. In this case, the decrement can simply act as if the increment occurred before it, and hence report that it did not bring the counter to zero. A decrement that races with a load must handle one additional case described in the next paragraph.

\myparagraph{Load} At first glace, the algorithm could try to just load the stored value, and return zero if the high bit is set. This however, is not necessarily correct if the stored value is zero. If the stored value is zero, the high bit might be about to be set, but an increment might race with it and bring the counter above zero. Reporting zero would therefore be incorrect. In order to achieve wait freedom, the load operation therefore attempts to help set the high bit. If it successfully sets the high bit, it can return zero. If it fails, the unsuccessful CAS will return the current value of the counter.

If the load operations successfully helps to store the high bit, one of the decrements still needs to take responsibility for being the one who brought the counter to zero. To achieve this, the helping operation additionally writes the second-highest bit, to indicate to the decrement operation that it was helped. If a decrement operation fails to CAS the high bit but detects the helper bit, it can then perform a fetch-and-store (exchange in C++) to remove the helper bit. If it removes the helper bit, it takes credit for bringing the counter to zero.

\begin{figure}
\begin{lstlisting}[basicstyle=\scriptsize\ttfamily,linewidth=\columnwidth]
@$\textbf{unsigned int}$@ zero = 1 << (b - 1);
@$\textbf{unsigned int}$@ help = 1 << (b - 2);

@$\textbf{unsigned int}$@ x;

bool @$\textbf{increment\_if\_not\_zero}$@() {
  auto val = x.fetch_add(1);
  return (val & zero) == 0; }
  
bool @$\textbf{decrement}$@() {
  if (x.fetch_sub(1) == 1) {
    @$\textbf{unsigned int}$@ e = 0;
    if (x.compare_exchange(e, zero)) return true;
    else if ((e & help) && (x.exchange(zero) & help)) return true;
  } return false; }
  
unsigned int @$\textbf{load}$@() {
  auto e = x.load();
  if (e == 0 && x.compare_exchange(e, zero | help)) return 0;
  return (e & zero) ? 0 : e; }
\end{lstlisting}
\caption{An implementation of a wait-free reference counter with constant time increment-if-not-zero, decrement, and load. Note that the \texttt{compare\_exchange} operation, if unsuccessful, atomically loads the value of \texttt{x} into \texttt{e}.}\label{fig:sticky-counter}
\end{figure}

\subsection{Primitives for weak reference counting}\label{sec:weak_ptr_primitives}

The addition of a weak reference count requires us to make changes to the use of the acquire-retire interface used behind our reference counting scheme. In the strong-only setting, a retired pointer always corresponds to a delayed decrement of the reference count. In the weak setting, our algorithm also needs to be able to delay decrements of the weak count.

Additionally, in the strong-only setting, obtaining a snapshot pointer to a managed object meant that the strong reference count was at least one, and since the pointer through which it was obtained is protected, it is guaranteed to remain at least one. However, this property cannot be guaranteed for a \emph{weak snapshot}, because a thread might be about to decrement the last remaining strong reference right as we acquire it. Therefore, to make weak snapshots safe, an additional round of deferral is required to defer the destruction of the managed object after its reference count hits zero. This guarantees that after an acquire, if the strong reference count is at least one, the object will not be destroyed until after the protection of the snapshot is released. We refer to the destruction of the managed object as a \emph{dispose} operation.

To facilitate these additional needs, instead of using a single instance of acquire-retire, our enhanced algorithm makes use of three instances---one for strong reference count decrements, one for weak decrements, and one for disposals.

\begin{figure}[t]
\begin{lstlisting}[basicstyle=\scriptsize\ttfamily,linewidth=\columnwidth,xleftmargin=5.0ex,numbers=left]
AcquireRetire<T> strongAR, weakAR, disposeAR;

void @$\textbf{delayed\_decrement}$@(T* p) {
  strongAR.retire(p);
  auto x = strongAR.eject();
  decrement(x); }
  
void @$\textbf{delayed\_weak\_decrement}$@(T* p) {
  weakAR.retire(p);
  auto x = weakAR.eject();
  weak_decrement(x); }
  
void @$\textbf{delayed\_dispose}$@(T* p) {
  disposeAR.retire(p);
  auto x = disposeAR.eject();
  dispose(x); }
  
T* @$\textbf{load\_and\_increment}$@(T** p) {
  auto ptr, guard = strongAR.acquire(p);
  if (ptr) increment(ptr);
  strongAR.release(guard);
  return ptr; }
  
T* @$\textbf{weak\_load\_and\_increment}$@(T** p) {
  auto ptr, guard = weakAR.acquire(p);
  if (ptr) weak_increment(ptr);
  weakAR.release(guard);
  return ptr; }
  
bool @$\textbf{increment}$@(T* p) {
  return p->ref_cnt.increment_if_not_zero(); }
  
void @$\textbf{weak\_increment}$@(T* p) {
  p->weak_cnt.increment_if_not_zero(); }
  
void @$\textbf{decrement}$@(T* p) {
  if (p->ref_cnt.decrement(1)) {
    delayed_dispose(p); } }

void @$\textbf{dispose}$@(T* p) {
  @$\textbf{destroy}$@(p->object);
  weak_decrement(p); }

void @$\textbf{weak\_decrement}$@(T* p) {
  if (p->weak_cnt.decrement(1)) {
    delete p; } }
    
bool @$\textbf{expired}$@(T* p) {
  return p->ref_cnt.load() == 0; }
\end{lstlisting}
\caption{Primitives for implementing deferred reference counting with support for weak pointers.}\label{fig:weak-ptr-primitives}
\end{figure}

Integrating these ideas, we extend the set of primitives for deferred reference counting with weak pointers as follows. Pseudocode is given in Figure~\ref{fig:weak-ptr-primitives}. The \textbf{delayed\_decrement}, \textbf{delayed\_weak\_decrement}, and \textbf{delayed\_dispose} operations make use of three different instances of acquire-retire to delay a decrement to the strong or weak reference count, or the destruction of the managed object, until it is no longer protected by a corresponding acquire.

\textbf{load\_and\_increment} and \textbf{weak\_load\_and\_increment} atomically load the value of the pointer stored at the given location and perform a safe increment of the strong or weak reference count respectively. Note that \texttt{load\_and\_increment} does not check whether the increment was successful, because these functions are only ever called on a pointer location that is storing a strong or weak reference respectively, and hence the reference count is already guaranteed to not be zero. It is a precondition violation to call this function on a pointer location that stores an object whose strong reference count is already zero.

\textbf{increment} and \textbf{weak\_increment} attempt to increment the reference count or weak reference count respectively. The first returns true if successful. Note that \texttt{weak\_increment} does not need to check for success, because objects with a zero weak reference count are instantly destroyed, and hence it would be unsafe to attempt to increment the counter anyway. \textbf{decrement} decrements the strong reference count, and if it reaches zero, queues up a delayed \emph{dispose}. A \textbf{dispose} destroys\footnote{We use destroy in the object-oriented sense to mean to recursively destroy all of its fields. If any of its fields are themselves reference-counted pointers, this would trigger their reference count decrements.} the managed object and decrements the weak reference count. Similarly, \textbf{weak\_decrement} decrements the weak reference count, and if it hits zero, immediately frees the managed object and its control data. Lastly, \textbf{expired} checks whether the managed object is still considered alive by checking that the reference count is not zero.

\subsection{Algorithms for atomic weak pointers}

Using the primitives from Figure~\ref{fig:weak-ptr-primitives}, the algorithms for storing and loading to/from and CASing into an atomic weak pointer are very similar to those in CDRC~\cite{anderson2021concurrent}. The main difference is that we must be careful to use the correct instance of acquire-retire for protection, and the correct kinds of increments/decrements. The algorithm that is most different from its strong counterpart is \texttt{get\_snapshot}. Pseudocode is given in Figure~\ref{fig:weak-ptr-implementation} and described below.

\begin{figure}[t]
\begin{lstlisting}[basicstyle=\scriptsize\ttfamily,linewidth=\columnwidth,xleftmargin=5.0ex,numbers=left]
void atomic_weak_ptr<T>::store(const weak_ptr<T>& desired) {
  if (desired.ptr) weak_increment(desired.ptr);
  auto old_ptr = this->ptr.exchange(desired.ptr);
  if (old_ptr) delayed_weak_decrement(old_ptr); }
  
weak_ptr<T> atomic_weak_ptr<T>::load() {
  auto ptr = weak_load_and_increment(addressof(this->ptr));
  return weak_ptr(ptr); }
  
bool atomic_weak_ptr<T>::compare_and_swap(
  const weak_ptr<T>& expected, const weak_ptr<T>& desired) {
  auto ptr, guard = weakAR.acquire(addressof(desired.ptr));
  if (compare_and_swap(this->ptr, expected.ptr, ptr)) {
    if (ptr) weak_increment(ptr);
    if (expected.ptr) delayed_weak_decrement(expected.ptr);
    weakAR.release(guard);
    return true; }
  else {
    weakAR.release(guard);
    return false; } }
    
weak_snapshot_ptr<T> atomic_weak_ptr<T>::get_snapshot() {
  while (true) {
    auto ptr, weak_guard = weakAR.acquire(addressof(this->ptr));
    auto _, dispose_guard = disposeAR.try_acquire(addressof(ptr));
    if (dispose_guard == @$\bot$@ && ptr) increment(ptr);
    
    if (ptr && !expired(ptr)) {
      weakAR.release(weak_guard);
      return weak_snapshot_ptr(ptr, dispose_guard); }
    else {
      disposeAR.release(dispose_guard);
      weakAR.release(weak_guard);
      if (ptr == null || this->ptr == ptr)
        return weak_snapshot_ptr(null); } }
        
void weak_snapshot_ptr<T>::release() {
  if (this->guard != @$\bot$@) disposeAR.release(this->guard);
  else decrement(this->ptr); }
\end{lstlisting}
\caption{C++-like pseudocode for operations on atomic weak pointers.}\label{fig:weak-ptr-implementation}
\end{figure}

\myparagraph{Storing a weak\_ptr in an atomic\_weak\_ptr} This works the same as storing a \texttt{shared\_ptr} in an \texttt{atomic\_shared\_ptr}. The algorithm increments the weak reference count of \texttt{desired}, uses a fetch-and-store (exchange in C++) to swap the managed object with the given one, and then performs a delayed decrement of the weak reference count of the previously stored object.

\myparagraph{Loading a weak\_ptr from an atomic\_weak\_ptr} This works the same as loading from an \texttt{atomic\_shared\_ptr}. The managed object is atomically loaded and has its weak reference count safely incremented, returning a \texttt{weak\_ptr} to the managed object.

\myparagraph{CASing into an atomic\_weak\_ptr} Compare and swap begins by protecting the pointer owned by \texttt{desired}. If the CAS is successful, it increments the weak reference count of desired and performs a delayed decrement of the weak reference count of \texttt{expected}. Note that the guard must be acquired before performing the CAS because otherwise, the CAS might succeed while another process clobbers \texttt{desired}, destroying it before the reference count increment happens.

\myparagraph{Creating a snapshot from an atomic\_weak\_ptr} Creating a snapshot from an \texttt{atomic\_weak\_ptr} is slightly more complicated than taking one from an \texttt{atomic\_shared\_ptr}. The main idea is to try to acquire a protected pointer to the managed object that prevents the object from being disposed, and, if the managed object has not expired (the strong reference count is at least one), return a snapshot containing the protected pointer. If the try\_acquire fails, the backup plan is to attempt to increment the reference count\footnote{This only happens with the hazard pointer implementation if too many snapshots are held at once such that the announcement array runs out of slots. EBR, IBR and Hyaline never fail.}. In case the managed object has already been disposed before protecting the pointer, the algorithm first acquires protection against a possible weak decrement, since, otherwise, the control data could be deleted mid-operation.

If the strong reference count is zero, the obvious algorithm would just return a snapshot containing a null pointer. However, this strategy would result in the operation not being linearizable, because the reference count could be in the process of being decremented right as the pointer is acquired. This would allow for situations where the \texttt{atomic\_weak\_ptr} always points to a live object, but the snapshot may return null if the object was replaced in between the acquire and the read of the reference count. Therefore, if the reference count is zero, the algorithm only returns a null pointer if the \texttt{atomic\_weak\_ptr} still manages the same acquired pointer. If not, the algorithm retries from the beginning.
\shepherd{Retrying in this manner causes \texttt{get\_snapshot} to be lock-free but not wait-free.}
 
\subsection{Example Usage}
\label{sec:weak-ptr-exp}

\shepherd{

An example of how to apply our \texttt{weak\_ptr} interface to Ramalhete and Correia's doubly-linked queue~\cite{doublelink} is shown in Figure~\ref{fig:weak-ptr-example}.
The \texttt{prev} pointer of each node is stored in an atomic weak pointer, whereas the \texttt{next} pointers are stored in atomic shared pointers.
The \texttt{critical\_section\_guard} (on lines~\ref{line:csguard1} and~\ref{line:csguard2}) is only needed if generalized acquire-retire was implemented from a protected-region SMR technique.
It is responsible for calling \texttt{begin\_critical\_section} in its constructor and \texttt{end\_critical\_section} in its destructor.
}

\begin{figure}[t]
  \begin{lstlisting}[basicstyle=\scriptsize\ttfamily,linewidth=\columnwidth,xleftmargin=5.0ex,numbers=left]
class doubly_linked_queue<V> {
  struct Node {
    V value;
    atomic_shared_ptr<Node> next;
    atomic_weak_ptr<Node> prev; 
    Node(V v) { value = v; next = null; prev = null; } };

  atomic_shared_ptr<Node> head, tail;

  void enqueue(V v) {
    shared_ptr<Node> new_node = shared_ptr<Node>::make_shared(v);
    critical_section_guard guard;     @\label{line:csguard1}@
    while (true) {
      snapshot_ptr<Node> ltail = tail.get_snapshot();
      new_node->prev.store(ltail);
      // Help the previous enqueue set its next ptr
      weak_snapshot_ptr<Node> lprev = ltail->prev.get_snapshot();
      if (lprev && lprev->next == null) lprev->next.store(ltail);
      if (tail.compare_and_swap(ltail, new_node)) {
        ltail->next.store(std::move(new_node));
        return; } } }

  std::optional<V> dequeue() {
    critical_section_guard guard;    @\label{line:csguard2}@
    while (true) {
      snapshot_ptr<Node> lhead = head.get_snapshot();
      snapshot_ptr<Node> lnext = lhead->next.get_snapshot();
      if (!lnext) return {};  // Queue is empty
      if (head.compare_and_swap(lhead, lnext)) {
        return {lnext->value}; } } } };
  \end{lstlisting}
  \caption{\shepherd{Ramalhete and Correia's concurrent doubly-linked queue~\cite{doublelink} implemented using our weak pointer interface (C++-like pseudocode).}}\label{fig:weak-ptr-example}
\end{figure}

%% file: experiments.tex

\section{Experimental Evaluation}
\label{sec:exp}

We implemented our techniques as a C++ library and evaluated them on a series of benchmarks. Our
experiments were run on a 4-socket 72-core machine ($4\times$ Intel(R) Xeon(R) E7-8867
v4, 2.4GHz) with $2$-way hyperthreading, a 45MB L3 cache, and 1TB of main memory. Memory
was interleaved across sockets using \emph{numactl -i all}, and we used the \emph{jemalloc}
allocator~\cite{jemalloc}. Experiments were written in C++ and compiled with GCC 9.2.1 with
\texttt{O3} optimization. Our experiments vary the number of threads from 1 to 192, which
allows us to measure the effect of oversubscription, as our hardware supports 144 threads.

\subsection{Comparison of manual and automatic techniques}

We applied the approach in Section~\ref{sec:automatic-smr} to three different manual SMR techniques, EBR~\cite{fraser2004practical}, IBR (more specifically, 2GEIBR) ~\cite{wen2018interval}, and Hyaline (more specifically, Hyaline-1)~\cite{nikolaev2019hyaline}, to construct three new concurrent reference counting implementations, which we call RCEBR, RCIBR, and RCHyaline, respectively.
The goal of this section is to understand the overhead of making manual techniques automatic as well as to compare the performance of RCEBR, RCIBR, and RCHyaline with the fastest existing reference counting algorithm.
The two fastest existing reference counting algorithms that we are aware of are FRC~\cite{tripp2018frc} and CDRC~\cite{anderson2021concurrent}.
We chose to compare with CDRC because FRC does not support marked pointers which are required in all of our benchmarks.
For consistency, we rename CDRC to RCHP in the graphs as it is a combination of hazard-pointers and reference counting.

As for manual techniques, we compare with HP, EBR, IBR, and Hyaline.
An important parameter to tune when using EBR and IBR is how often the global epoch gets incremented.
Incrementing too often could bottleneck scalability whereas incrementing infrequently would increase memory usage.
For EBR and RCEBR, we found the ideal rate to be one increment every ~10 allocations and for IBR and RCIBR, we found this to be one increment every ~40 allocations.

For both HP and RCHP, we found that prefetching appropriately significantly increased throughput. In particular, before announcing a pointer in the hazard array, we prefetch the cache line that it points to because there is a good chance we will deference the pointer after succeeding in announcing it.
The benefit of this is that we can start loading the cache line before the memory barrier, \Hao{People might not know what fencing means. Maybe call it a memory barrier instead} which is an expensive operation. Note that due to this prefetching optimization, our throughput reported here for HP and RCHP is greater than the throughput of the same schemes in CDRC~\cite{anderson2021concurrent}.

To benchmark performance, we applied these memory reclamation techniques to three different lock-free data structures: Harris-Michael list ~\cite{harris2001list, michael2002high}, Michael hash table~\cite{michael2002high}, and Natarajan-Mittal tree~\cite{natarajan2014fast}. 

It has been noted that HP and IBR are not safe to use with the Natarajan-Mittal tree directly~\cite{anderson2021concurrent}.
This is because traversals in the Natarajan-Mittal tree can continue through marked nodes.
We still include these numbers in our experiments for reference, even though these experiments occasionally crash.
Modifying the Natarajan-Mittal tree to work with HP and IBR would likely make it slower.
Note that an advantage of RCHP and RCIBR is that they work with Natarajan-Mittal tree without any such modifications.

\begin{figure}
 	\centering
 	\includegraphics[width=0.49\textwidth]{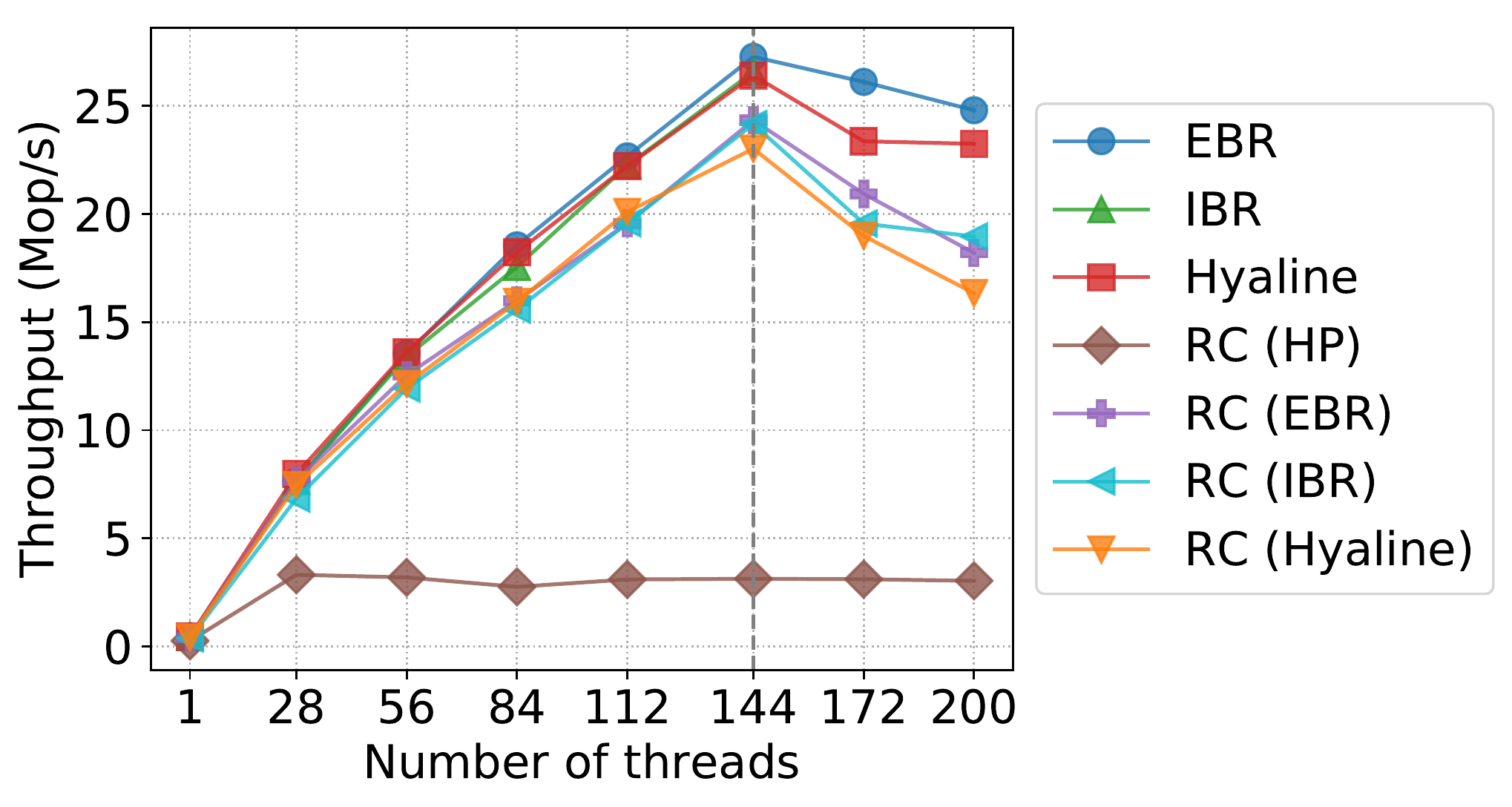}
 	\caption{Natarajan-Mittal tree - Range query experiments: 50\% updates, 50\% range queries of size 64.}
 	\label{fig:smr-exp-rq}
 \end{figure}

\textbf{Range query workload.} We begin by analyzing the experiment shown in Figure~\ref{fig:smr-exp-rq}. In this workload, we initialized the Natarajan-Mittal tree with 100K keys randomly selected from the key range $[0, 200K)$, and then performed update operations (half insert, half delete) and range queries\footnote{We use a sequential range query algorithm, which is not linearizable.} with equal probability. 
Each update operation selects a uniform random key from $[0, 200K)$ to insert/delete and each range query selects a uniform random key $k$ from the same range and queries for all keys in the interval $[k, k+64)$.
In this experiment, we found that RCEBR, RCIBR, and RCHyaline outperform RCHP by more than 7x on 144 threads.
This is because during a range query, the entire path from the current node to the root needs to be protected by \op{snapshot\_ptr}s, so RCHP eventually runs out of announcement locations and starts relying on reference count increments, which is significantly more expensive.
RCEBR, RCIBR, and RCHyaline also performs similarly to their manual counterparts, performing within 10-15\% at 144 threads.
 
\textbf{Other workloads.} Figure~\ref{fig:smr-exp} shows the throughput and memory usage of these SMR technique on a wide variety of workloads.
These workloads only contain updates and single point lookups.
For example, Figure~\ref{subfig:bst-100K-10} shows a workload where the Natarajan-Mittal tree is initialized with 100K keys, and each process performs $10\%$ update operations and $90\%$ lookups.
Again, all keys are chosen uniformly randomly from a key range twice the initial size of the data structure.
For the hash table experiments, we initialized the number of buckets so that the average load factor is 1.

When update frequency is low (Figure~\ref{subfig:bst-100K-1}), RCEBR has almost the exact same throughput as EBR and RCHyaline is actually slightly faster than Hyaline.
However, RCIBR ends up being about 20\% slower than IBR and this overhead comes from two main factors.
First, RCIBR adds both a reference count and a birth epoch to each node, and this increase in size accounts for about half of the performance difference.
Second, each \op{try\_acquire} in RCIBR requires reading a thread local variable storing the process id and this access is surprisingly slow, accounting for the other half of the performance difference.
Overall, on the BST experiments with 144 threads, RCEBR performs within 10\% of EBR (in terms of throughput) and RCHyaline performs within 15\% of Hyaline.
Also, RCEBR is up to 1.7x faster than RCHP in Figure~\ref{subfig:bst-100K-10}.

In the non-oversubscribed scenarios, the automatic version of each memory reclamation scheme tends to use a similar amount of memory to the manual version. 
However in the linked list experiment and also in oversubscribed cases, the automatic version tends to have several times more memory overhead.
This is because in the reference counting techniques each retired pointer could recursively prevent the collection of many nodes beyond the one it directly points to.

\Hao{explain why the time overhead of reference counting is higher in the linked list. Space overhead seems to get pretty high as well (something like 2-4x more space usage).}

\subsection{Evaluation of atomic weak pointers}

We compare our implementation of atomic weak pointers with the best known existing lock-free implementation, the just::thread library~\cite{just19}, and against a manually memory-managed data structure. For our comparison we use the doubly linked queue of Ramalhete and Correia~\cite{doublelink}. This queue is a good candidate since it uses back pointers that can be represented using weak pointers. For this comparison, we use our reference counting library powered by the hazard pointer implementation of acquire-retire. We found that the main bottleneck of the throughput of the data structure is the contention on the CAS operations, and hence the different choices of acquire-retire implementation only made minor differences to the performance.

The original implementation of the data structure does not use a general purpose memory management scheme, but actually uses a customized version of hazard pointers specifically engineered for it. This modified hazard pointers scheme allows announced nodes to protect not only themselves, but also the nodes adjacent to them. This reduces the number of memory fences required by the algorithm. For this reason, it is not likely that a general purpose memory management scheme would outperform it.

\begin{figure}[t]
 \centering
 \includegraphics[width=0.8\columnwidth]{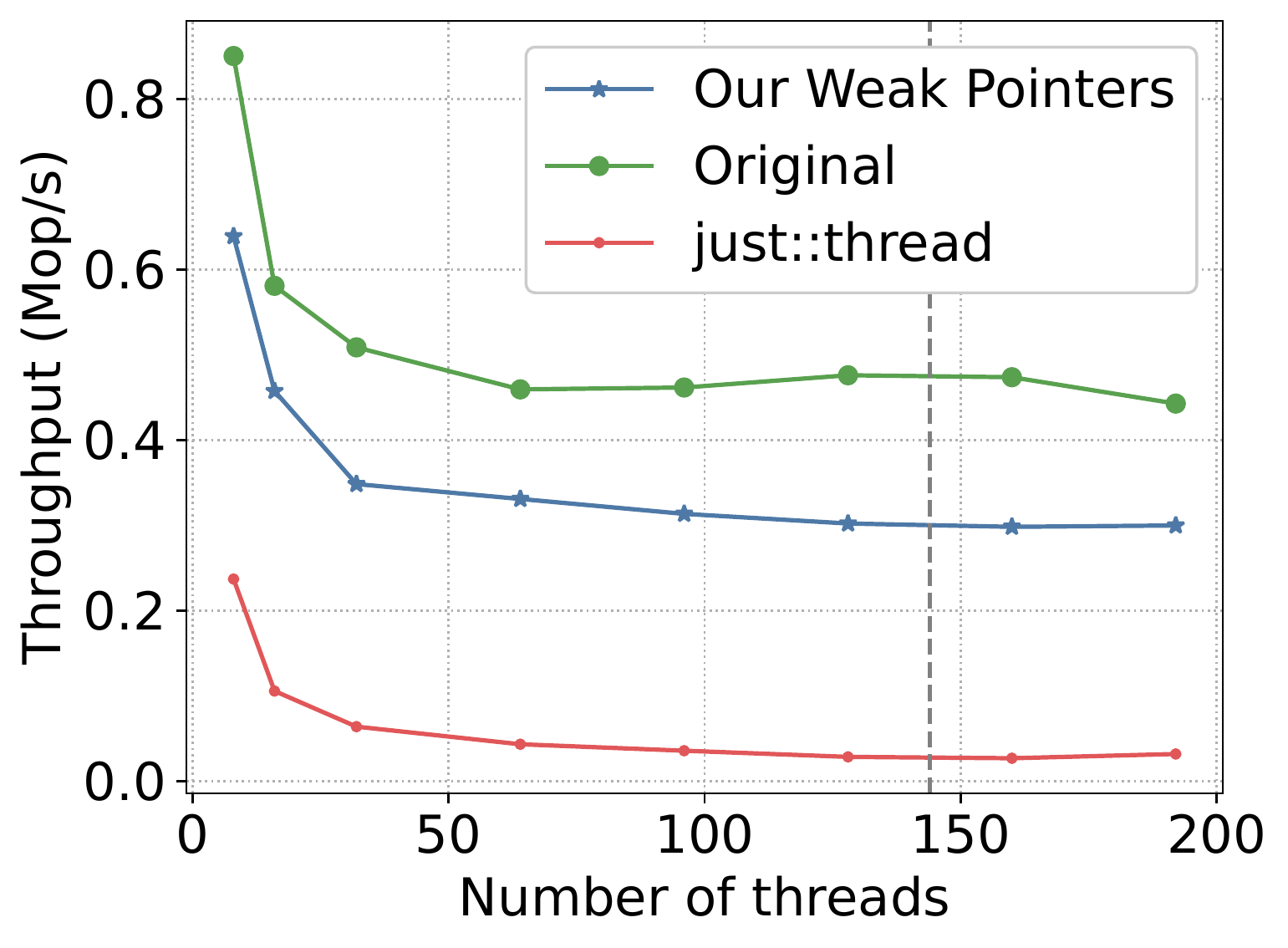}
 \caption{Benchmark results for atomic weak pointers. Original is the optimized doubly linked queue of Ramalhete and Correia~\cite{doublelink} that uses a custom manual memory management technique. Our algorithm uses atomic weak pointers powered by the hazard pointer implementation of acquire-retire. \texttt{just::thread} is a commercial library of atomic shared and weak pointers. \Hao{update caption}}\label{fig:weak-ptr-experiments}
\end{figure}

In our experiment, we initialize a single queue with $P$ elements, and have $P$ threads each thread repeatedly pop an element from the queue and then reinsert it. We then measure the number of such operations that were performed per second. Each benchmark is repeated five times for stability. The results of this experiment are depicted in Figure~\ref{fig:weak-ptr-experiments}.

The biggest difference in performance occurs at $P = 1$ (not depicted on the plot due to scale), where the original implementation is 4.5x faster than our weak pointers, and 67x faster than just::thread. At $P = 8$ threads, our weak pointer implementation is just $19\%$ slower than the manual approach, and $4.2$x faster than just::thread. This trend roughly continues to $P = 192$, where our weak pointers are $33\%$ slower than the manual approach, but 10x faster than just::thread. Given that the original implementation uses a memory management approach that is both manual and customized to the data structure at hand, these results are very promising for a completely automatic approach. Furthermore, we substantially outperform the best existing automatic approach at all thread counts.

\input{smr-benchmarks}

%% file: smr-benchmarks.tex
\begin{figure*}
  \centering

    \begin{subfigure}{\textwidth}
      \centering
  		\includegraphics[width=0.8\textwidth]{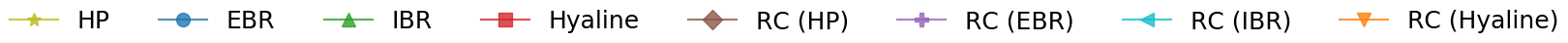}
    \end{subfigure}

  \begin{subfigure}{0.49\textwidth}
    \centering
    \includegraphics[height=3.1cm]{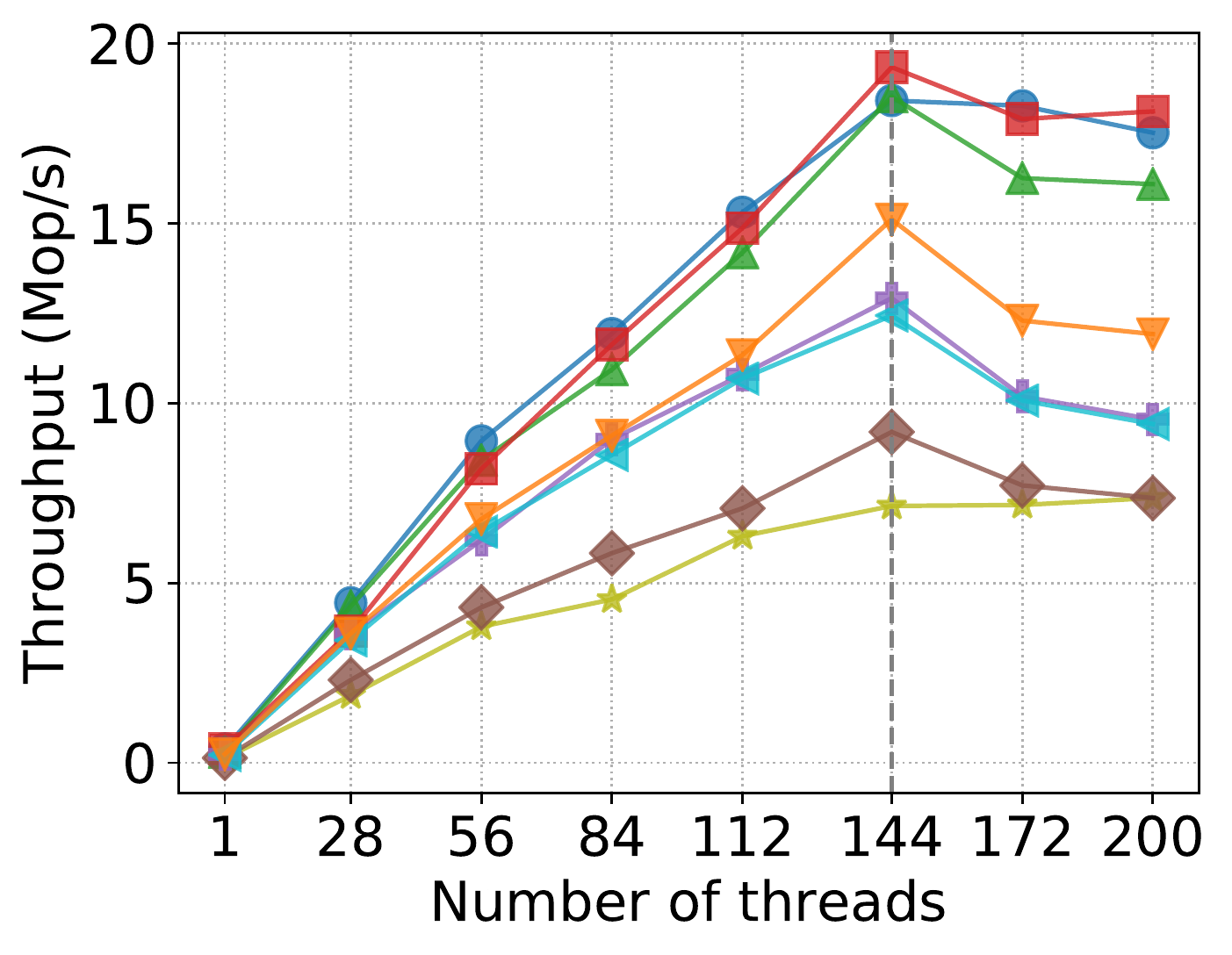}
    \includegraphics[height=3.1cm]{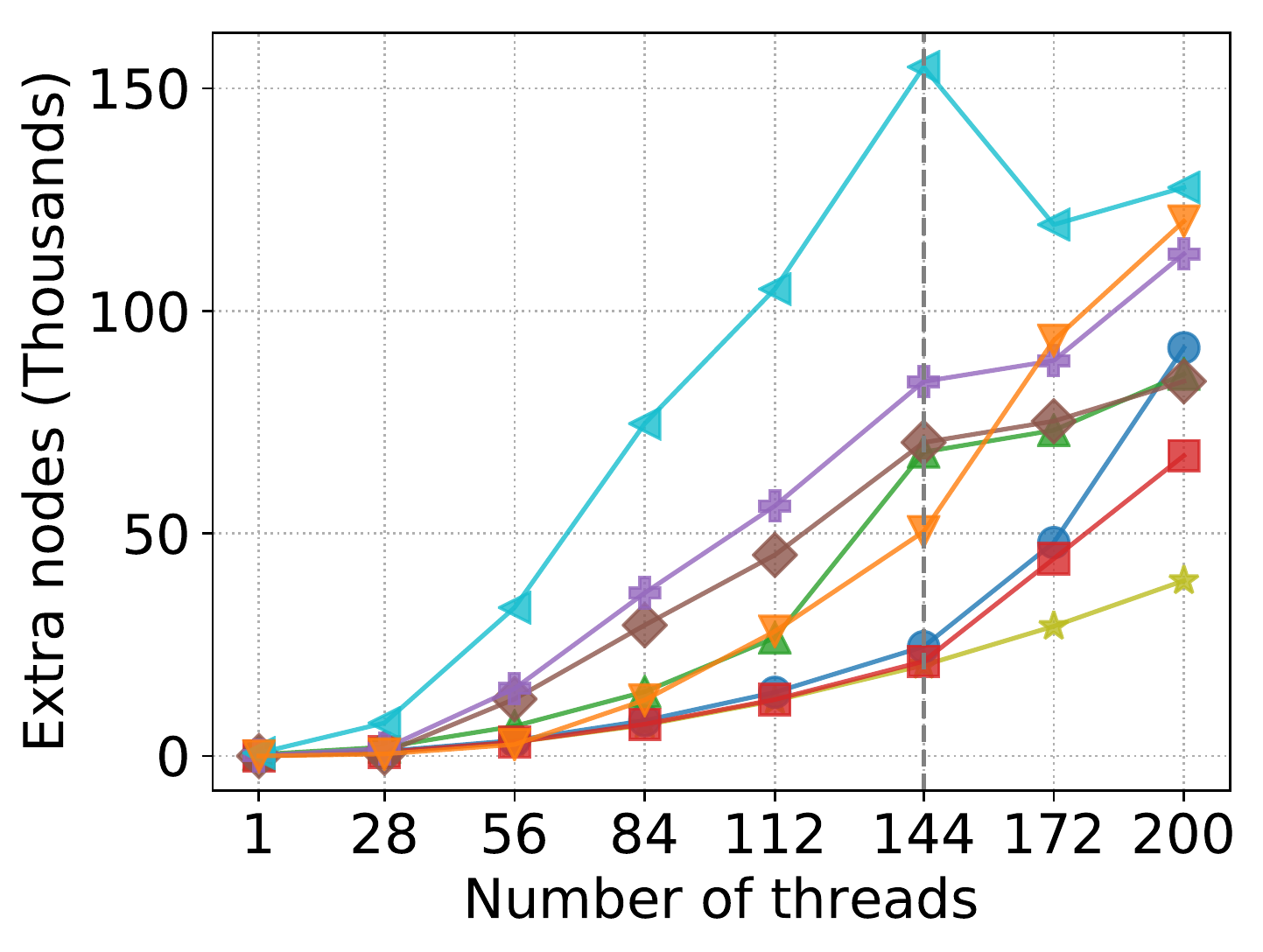}
    \captionsetup{justification=centering}
    \caption{List. N=1000, updates=10\%. Throughput (L), Memory (R)}\label{subfig:list-1000-10}
  \end{subfigure}
  \begin{subfigure}{0.49\textwidth}
    \centering
    \includegraphics[height=3.1cm]{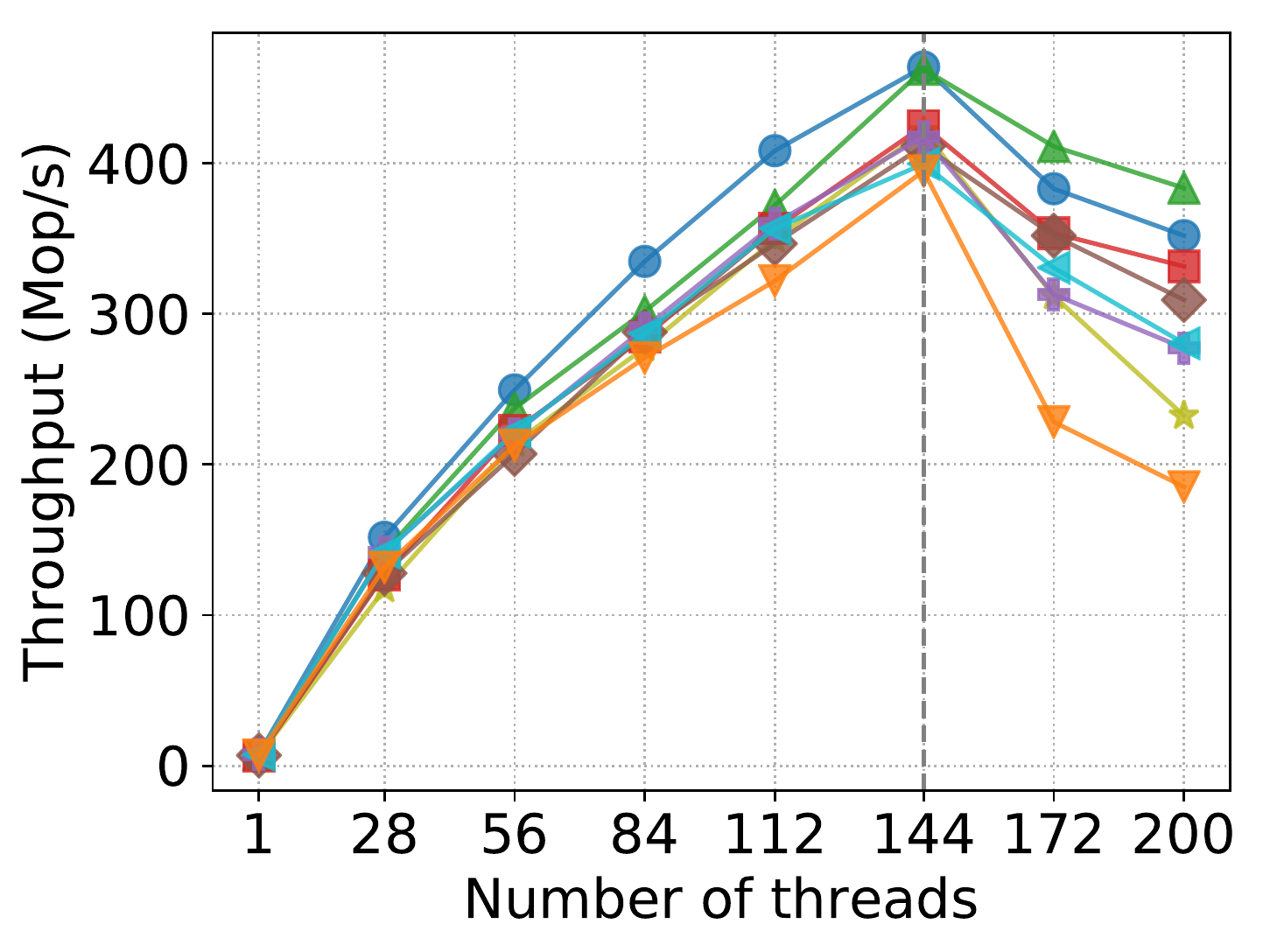}
    \includegraphics[height=3.1cm]{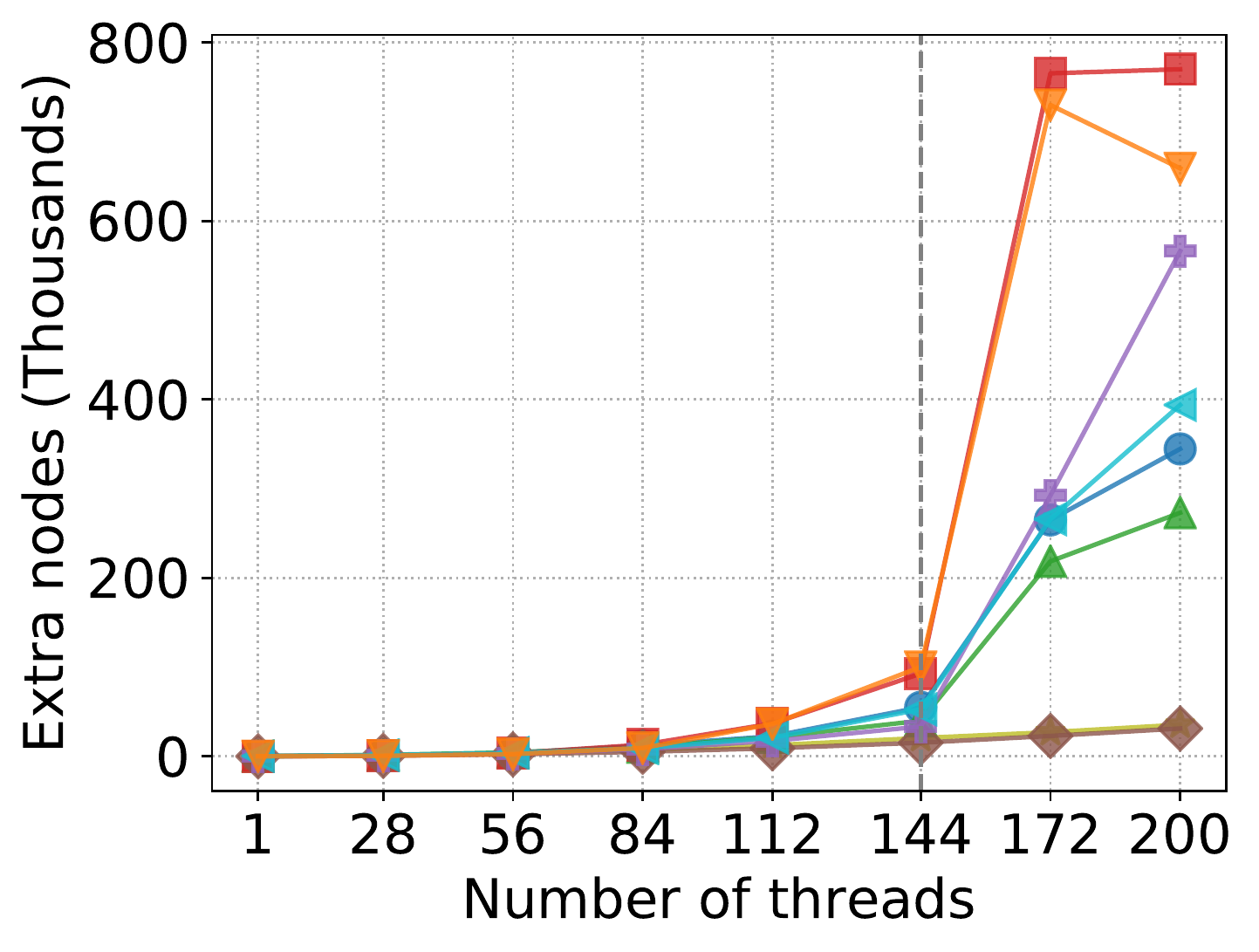}
    \captionsetup{justification=centering}
    \caption{Hashtable. N=100K, updates=10\%. Throughput (L), Memory (R)}\label{subfig:table-100K-10}
  \end{subfigure}
  
    \begin{subfigure}{0.49\textwidth}
        \centering
        \includegraphics[height=3.1cm]{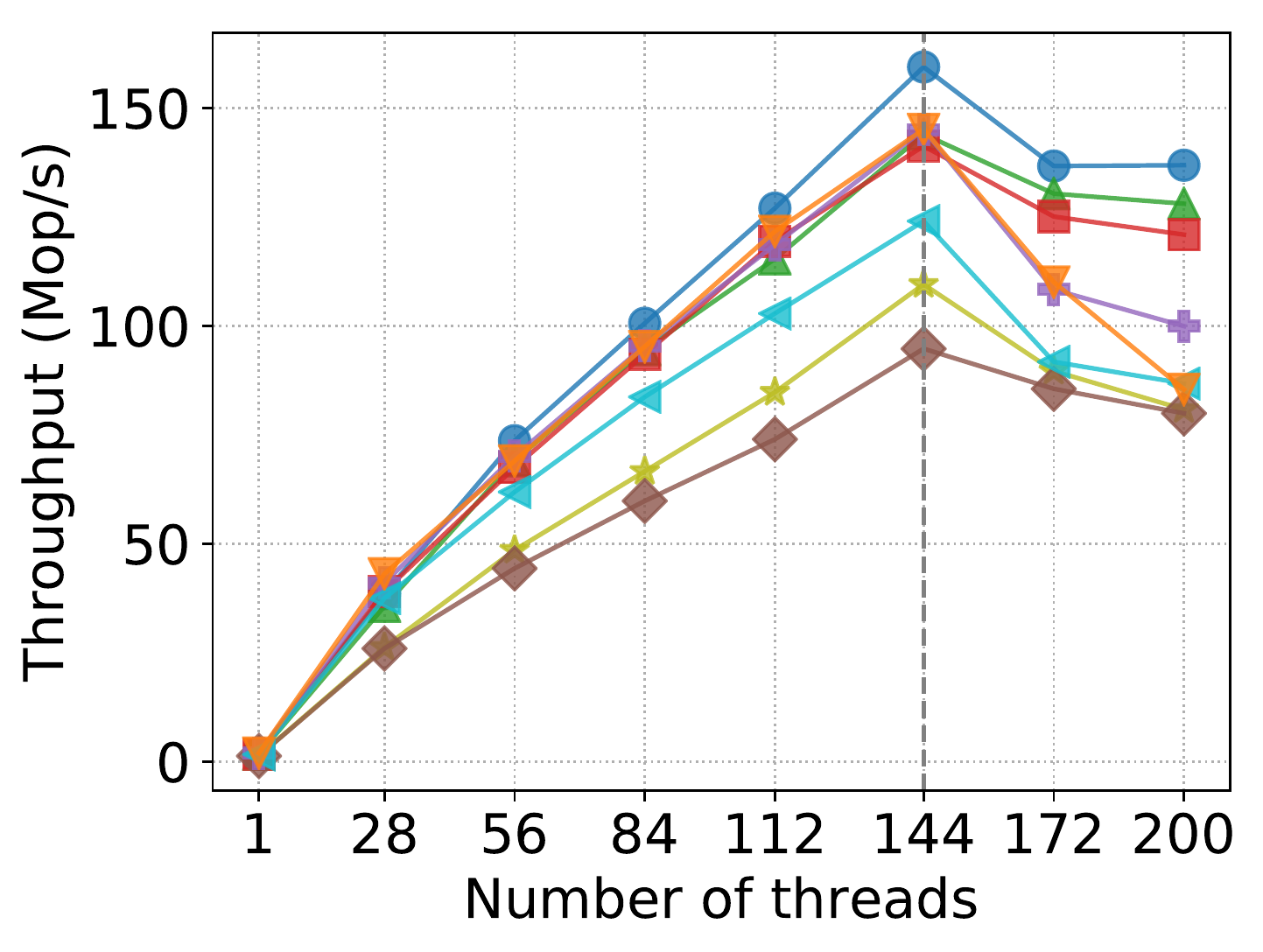}
        \includegraphics[height=3.1cm]{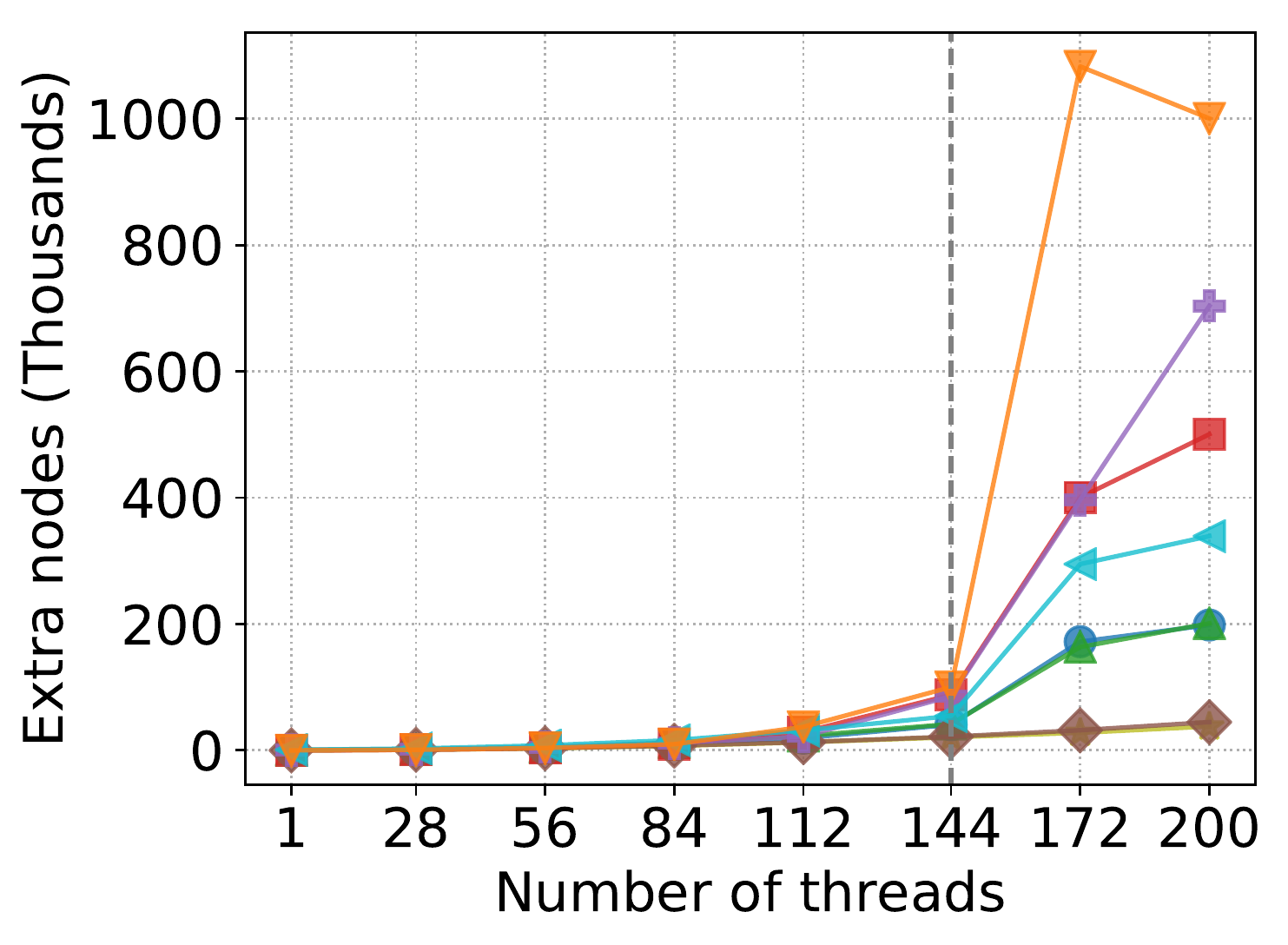}
        \captionsetup{justification=centering}
        \caption{BST. N=100K, updates=10\%. Throughput (L), Memory (R)}\label{subfig:bst-100K-10}
      \end{subfigure}
      \begin{subfigure}{0.49\textwidth}
        \centering
        \includegraphics[height=3.1cm]{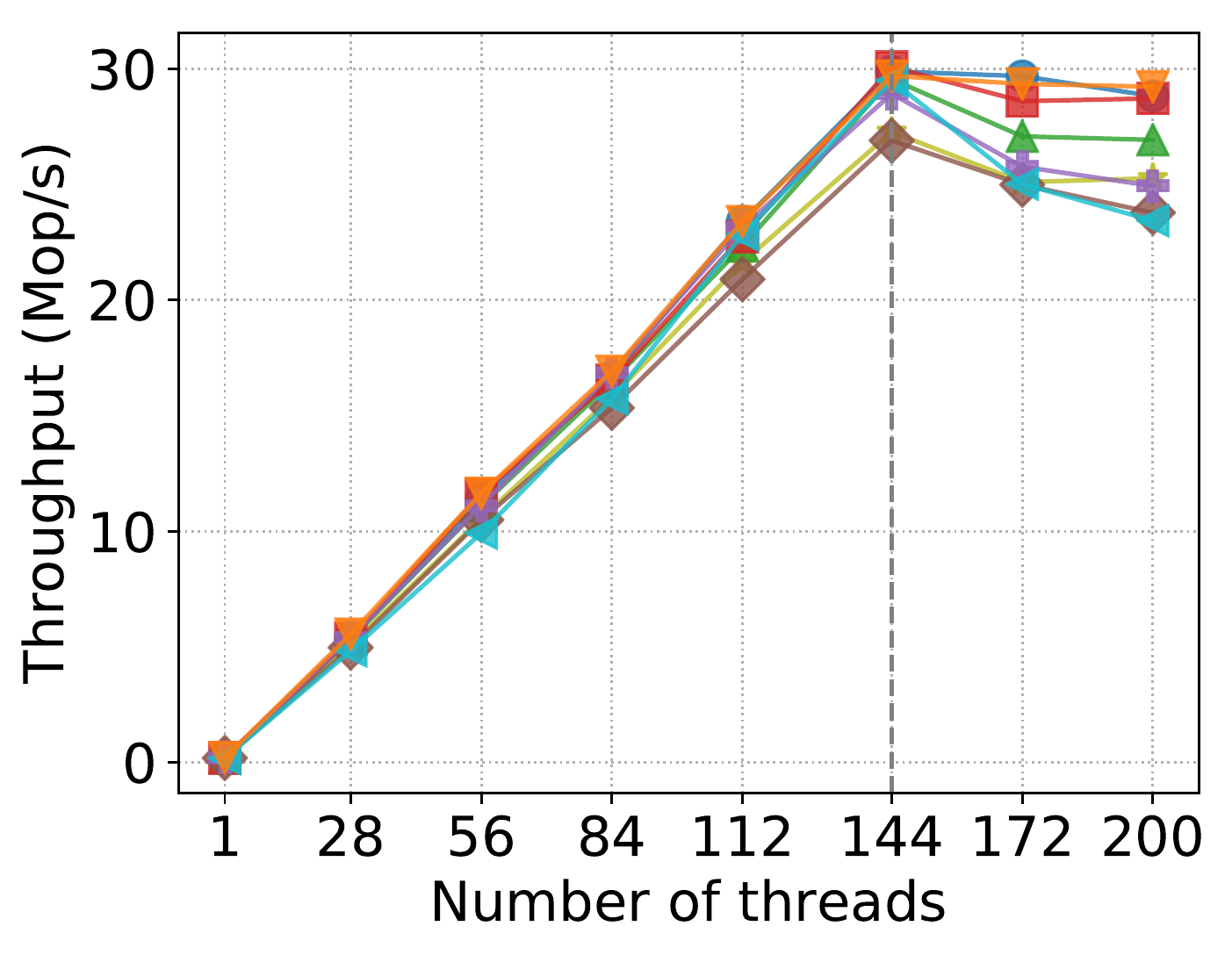}
        \includegraphics[height=3.1cm]{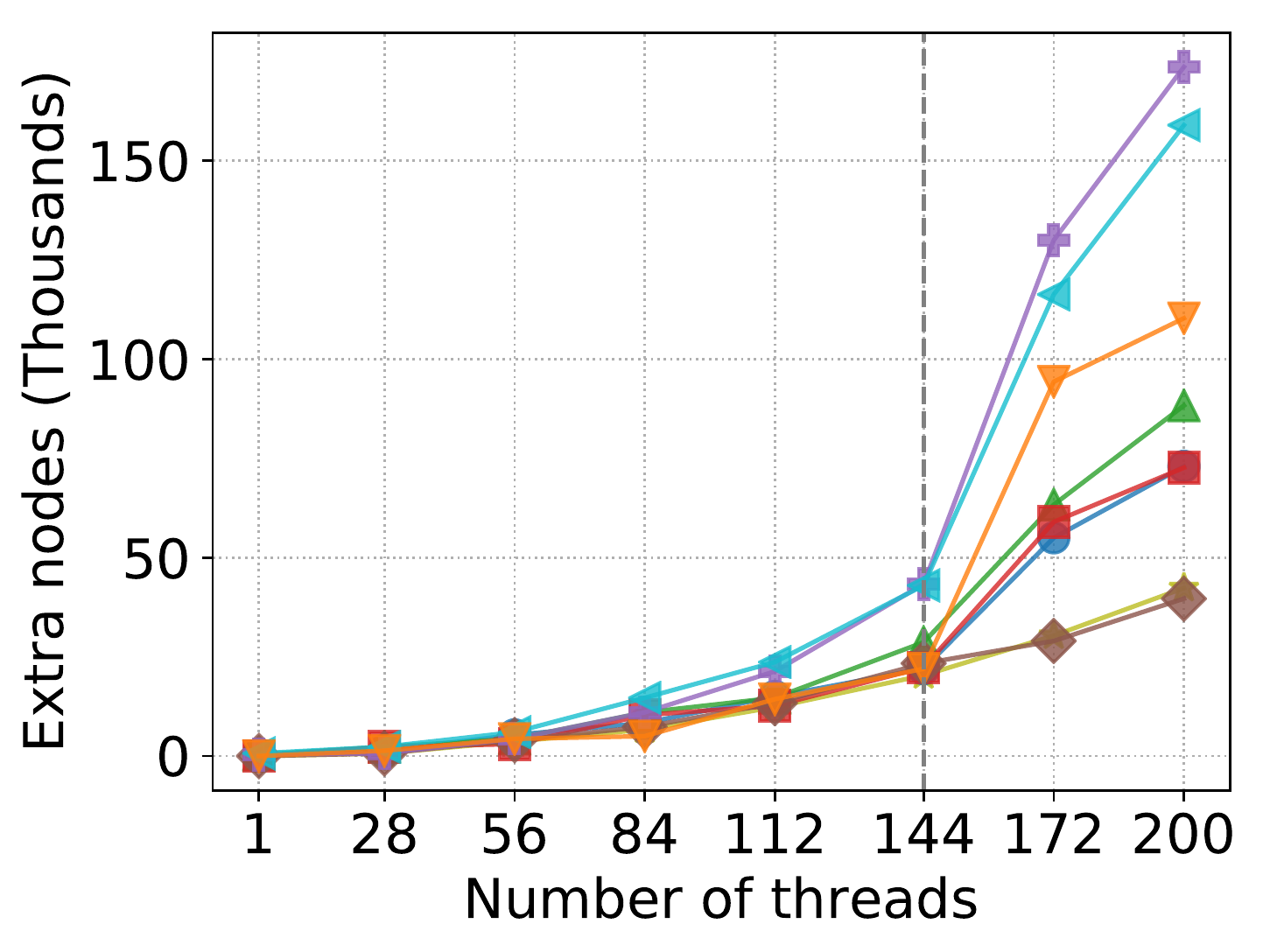}
        \captionsetup{justification=centering}
        \caption{BST. N=100M, updates=10\%. Throughput (L), Memory (R)}\label{subfig:bst-100M-10}
      \end{subfigure}
  
  \begin{subfigure}{0.49\textwidth}
      \centering
      \includegraphics[height=3.1cm]{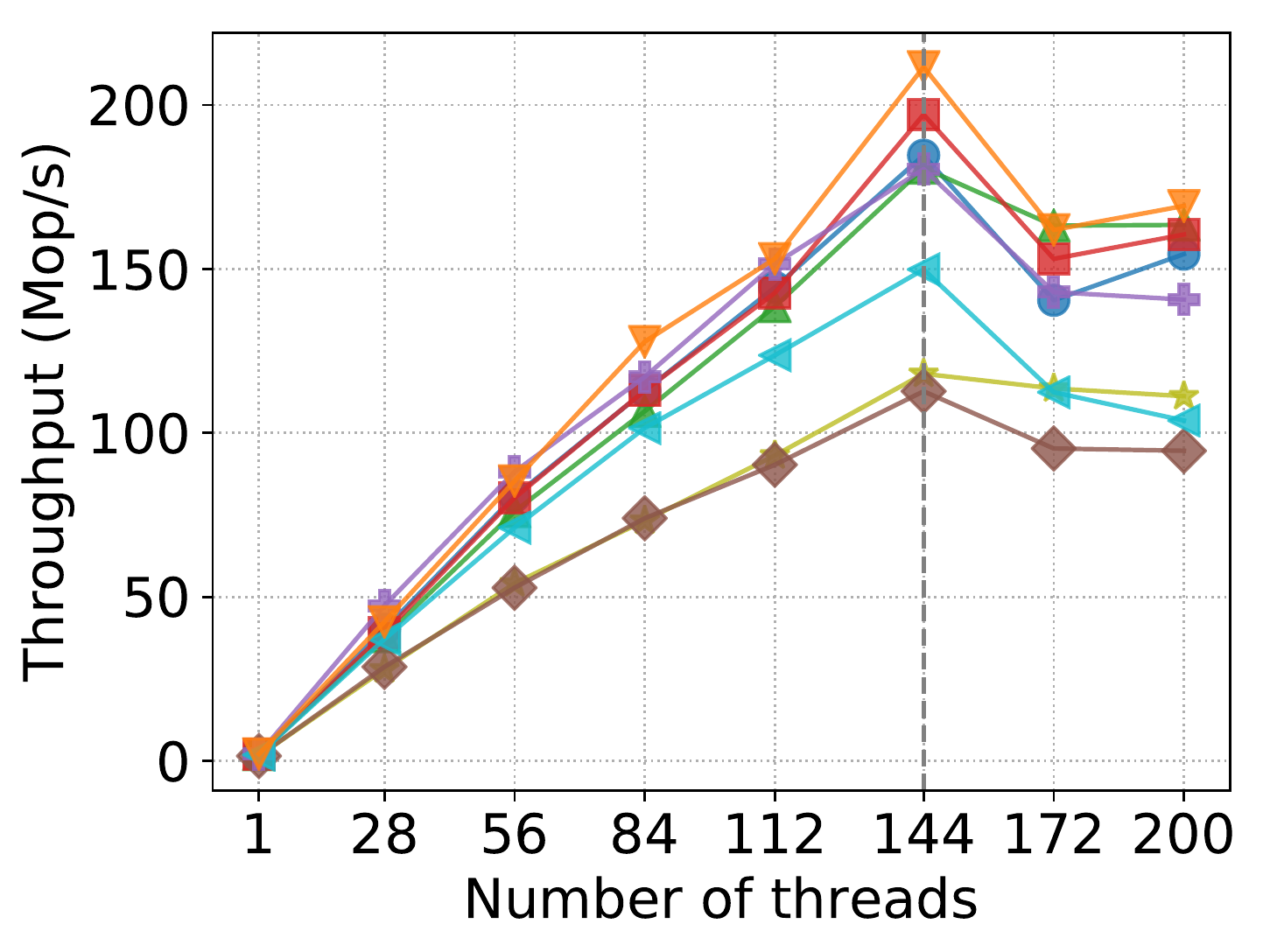}
      \includegraphics[height=3.1cm]{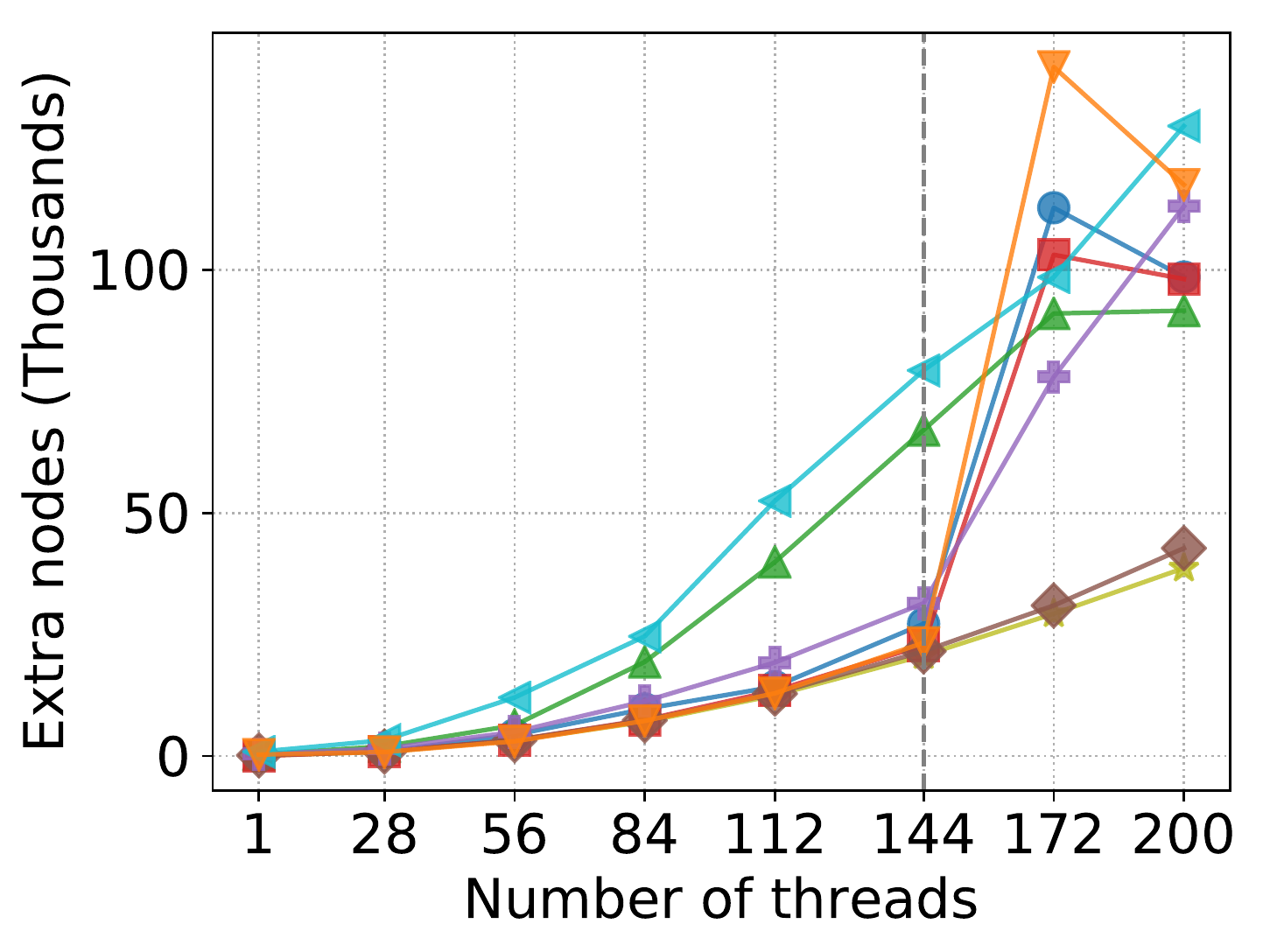}
      \captionsetup{justification=centering}
      \caption{BST. N=100K, updates=1\%. Throughput (L), Memory (R)}\label{subfig:bst-100K-1}
    \end{subfigure}
    \begin{subfigure}{0.49\textwidth}
      \centering
      \includegraphics[height=3.1cm]{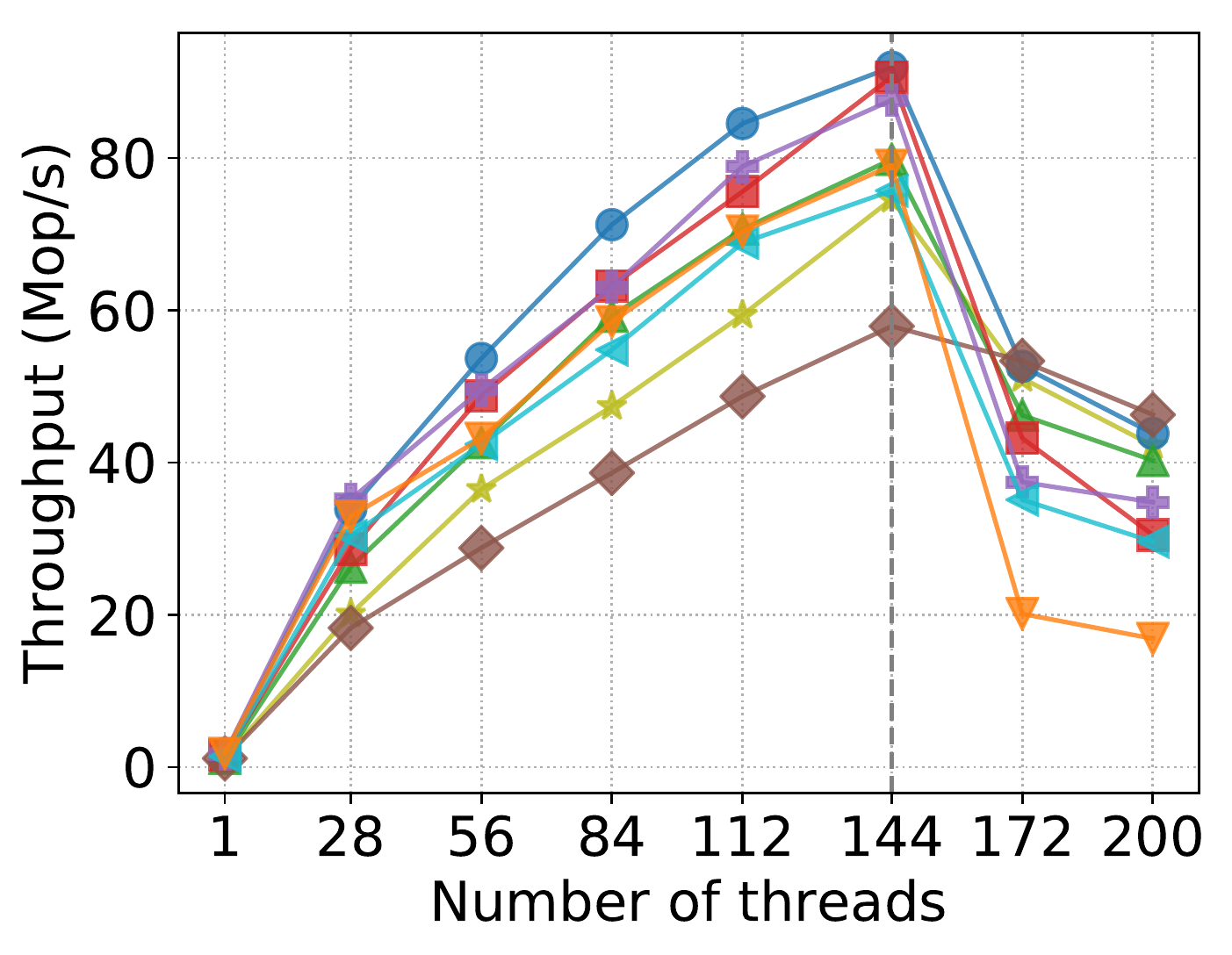}
      \includegraphics[height=3.1cm]{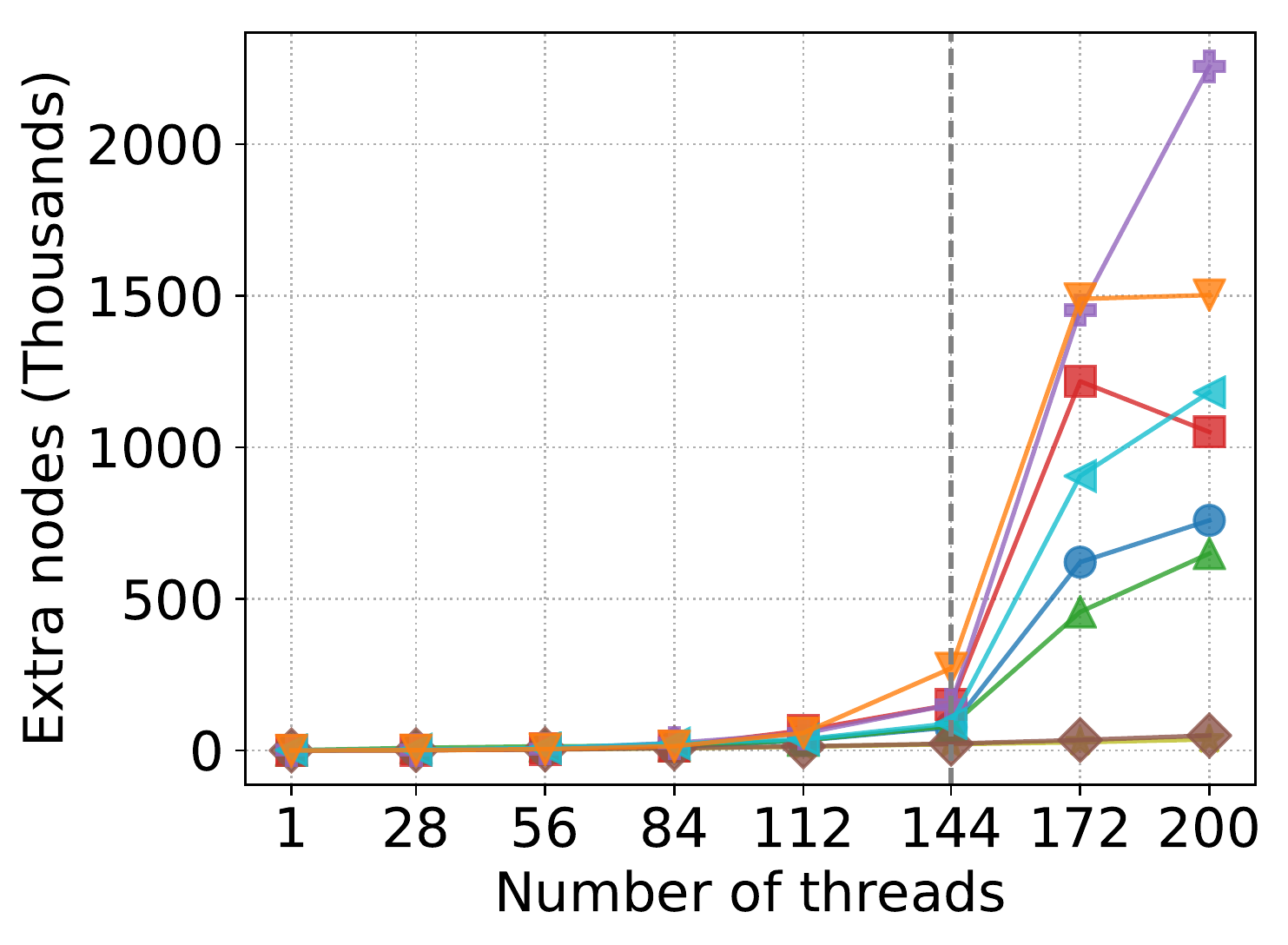}
      \captionsetup{justification=centering}
      \caption{BST. N=100K, updates=50\%. Throughput (L), Memory (R)}\label{subfig:bst-100K-50}
    \end{subfigure}
  
  \caption{Benchmark comparing manual and automatic SMR techniques. Figure~\ref{subfig:list-1000-10} shows results for a Harris-Michael list, Figure~\ref{subfig:table-100K-10} for a Michael hash table, and Figures~\ref{subfig:bst-100K-10}--\ref{subfig:bst-100K-50} for various workloads on a Natarajan-Mittal tree.}
  \label{fig:smr-exp}
\end{figure*}

%% file: related.tex

\section{Related Work}
\label{sec:related}

\subsection{Manual SMR}

Techniques for manual SMR typically fall broadly into one of two categories, protected-pointer-based methods or
protected-region-based methods.

\myparagraph{Protected-pointer-based methods}
Protected-pointer-based methods typically work by identifying specific objects/memory locations that are currently in use and hence
should not be destroyed/freed. The collection part of the algorithm is responsible for ensuring that it never frees something that
is currently indicated as being in use. Hazard pointers~\cite{michael2004hazard} is one of the most widely used protected-pointer-based
techniques. The main idea is that every process has some globally visible array of ``hazard pointers''. When a process wishes to read
a mutable shared pointer, it \emph{announces} its intention to do so by writing the stored pointer into one of the hazard pointers.
This may require a retry if the value of the stored pointer changes before the announcement is complete. When the process has finished
reading or manipulating the shared object, it \emph{releases} the hazard pointer by clearing the announcement. When a process removes
a node from the data structure and wishes to free it, it instead \emph{retires} the node, which places it in a \emph{retired list} of
nodes that are pending deletion.
A process that wishes to reclaim memory must scan the hazard array of every process to ensure that it does not reclaim anything protected
by a hazard pointer. Nodes in the retired list that are not protected are safe to free.

Several variants of hazard pointers exist, many of them designed to help implement other memory reclamation schemes.
Herlihy {\it et al.}~\cite{Herlihy05} develop Pass The Buck (PTB), which is used to implement their algorithm for lock-free reference
counting. Correia {\it et al.}~\cite{correia2021orcgc} develop pass-the-pointer (PTP), which improves on the memory bounds of
traditional hazard pointers and is used to implement their own lock-free reference counting algorithm, OrcGC. Anderson {\it et al.}~\cite{anderson2021concurrent}
introduce acquire-retire, the first constant-time implementation of hazard pointers, which also supports the additional property of
allowing a pointer to be retired multiple times concurrently. acquire-retire is used to implement CDRC.

\myparagraph{Protected-region-based methods}
Rather than protecting specific objects/memory locations, protected-region-based methods protect groups of objects. This
generally results in lower synchronization cost (fewer memory barriers) and hence higher throughput, but at the cost of wasting
more memory, since many objects will be protected even when they do not need to be. Epoch-based reclamation (EBR)~\cite{fraser2004practical}
and Read-copy-update (RCU)~\cite{mckenney2008rcu}
are the most widely used protected-region-based techniques. In EBR, the algorithm maintains a global timestamp called the epoch. Whenever
a memory location is retired, it is placed in a retired list corresponding to the current epoch. When the user wishes to begin an operation
that will access or modify shared state, the executing thread announces the value of the current epoch. When every thread has announced the
value of the current epoch, the retired list from the previous epoch can be freed and the epoch can advance to the next value. Note that
this is safe because if an object is retired at epoch $e$ and every process has subsequently announced epoch $e+1$, then any thread that
was performing an operation at the time of the retire has since completed.
DEBRA~\cite{brown2015reclaim} is an optimized implementation of EBR with better practical performance.

Hazard Eras (HE)~\cite{ramalhete2017brief,nikolaev2020universal} is a combination of protected-pointer- and
protected-region-based methods. In HE, acquired pointers don't announce the pointer itself, but rather the epoch on which it was read. If
the epoch changes infrequently, this results in fewer memory barriers than a full-blown protected-pointer-based scheme. In HE and
Interval-based Reclamation (IBR)~\cite{wen2018interval}, each allocated object is tagged with a birth epoch. In IBR, a retired object is safe
to reclaim when no announced epoch intersects its birth-death interval.

Hyaline~\cite{nikolaev2019hyaline,nikolaev2021snapshot} is a variant of EBR that tags each retired object with a counter corresponding to
the number of active operations. When an operation completes, it can decrement one from every object that retired during it. The operation
that brings a counter to zero is responsible for freeing it. Crystalline~\cite{nikolaev2021brief} extends Hyaline with wait freedom.

\subsection{Lock-free Reference Counting}

Lock-free reference counting (LFRC) was first described by Detlefs {\it et al.}~\cite{DMMS02}, but their algorithm requires a DCAS operation
(a CAS on two independent words), which is not supported by any current architecture. Herlihy {\it et al.}~\cite{Herlihy05} use their
PTB technique to obtain an algorithm for single-word lock-free reference counting (SLFRC). The idea is to use PTB to protect the control
block of the object (the area of memory containing the reference count) from unsafe reclamation while a process is attempting to increment
the reference count. To avoid racing on the reference count, a CAS loop is used for increments, such that if the reference count reaches zero,
the operation retries on the new stored value. Sundell~\cite{Sundell05} developed the first wait-free algorithm for reference counting, however,
some of their operations cost $O(P)$ time, which is expensive in practice.

The split reference count technique~\cite{williams2012book} is a non-SMR-based lock-free solution for atomic reference counting. It involves
splitting the reference count into an internal count, and an external count on each mutable shared reference. Loads from shared references
increment the corresponding external count, while local releases decrement the internal count instead. When a shared reference is discarded,
its accumulated external count minus one is added to the internal count. While this technique is appealing in its lack of reliance on SMR,
it tends to scale poorly in practice since loads must be performed with a double-word CAS to increment the external count.

The major performance drawback of reference counting is the necessity to increment the reference count each time an object is read. Recent work has addressed this by
developing solutions for reference counting that allow safe reads without incrementing the reference count. Tripp {\it et al.}~\cite{tripp2018frc} implement
Fast Reference Counter (FRC). FRC uses deferred reference counting and a per-thread root set (equivalent to an announcement array of hazard pointers)
to achieve low contention and enable safe reads of managed objects without incrementing the reference count. Correia {\it et al.}~\cite{correia2021orcgc}
develop OrcGC, which uses their PTP technique to implement reference-counted pointers that can also be safely read without increasing the reference count. Finally,
Anderson {\it et al.}~\cite{anderson2021concurrent} develop Concurrent Deferred Reference Counting (CDRC), which uses the acquire-retire technique to defer
reference count decrements and also enable safe reads without incrementing the reference count.

%% file: conclusion.tex

\section{Conclusion}
\label{sec:conclusion}

In this work, we showed that an automatic memory reclamation technique can compete with the best manual techniques, and showed that such a technique can also
support atomic weak pointers. Though perhaps it is not yet time to completely retire manual memory reclamation, we believe that these results show, even stronger
than previous results, that we are getting close, and that automatic memory management should be preferable in a majority of situations.
